\documentclass[multphys,vecphys]{svmult}

\usepackage{makeidx}     % allows index generation
\usepackage{graphicx}    % standard LaTeX graphics tool
\usepackage{multicol}    % used for the two-column index
\usepackage{hyperref}
\usepackage{amssymb}
\makeindex             % used for the subject index
\begin{document}

\title*{Quantum phases and phase transitions\\ of Mott insulators}
\author{Subir Sachdev}
\institute{Department of Physics, Yale University, P.O. Box
208120,\\ New Haven CT 06520-8120, USA,~
\texttt{subir.sachdev@yale.edu} }
\maketitle
\begin{abstract}
This article contains a theoretical overview of the physical
properties of antiferromagnetic Mott insulators in spatial
dimensions greater than one. Many such materials have been
experimentally studied in the past decade and a half, and we make
contact with these studies. The simplest class of Mott insulators
have an even number of $S=1/2$ spins per unit cell, and these can
be described with quantitative accuracy by the bond operator
method: we discuss their spin gap and magnetically ordered states,
and the transitions between them driven by pressure or an applied
magnetic field. The case of an odd number of $S=1/2$ spins per
unit cell is more subtle: here the spin gap state can
spontaneously develop bond order (so the ground state again has an
even number of $S=1/2$ spins per unit cell), and/or acquire
topological order and fractionalized excitations. We describe the
conditions under which such spin gap states can form, and survey
recent theories (T. Senthil {\em et al.}, cond-mat/0312617)
of the quantum phase transitions among these
states and magnetically ordered states. We describe the breakdown
of the Landau-Ginzburg-Wilson paradigm at these quantum critical
points, accompanied by the appearance of emergent gauge
excitations.
\end{abstract}

\section{Introduction}
\label{sec:intro}

The physics of Mott insulators in two and higher dimensions has
enjoyed much attention since the discovery of cuprate
superconductors. While a quantitative synthesis of theory and
experiment in the superconducting materials remains elusive, much
progress has been made in describing a number of antiferromagnetic
Mott insulators. A number of such insulators have been studied
extensively in the past decade, with a few prominent examples
being CaV$_4$O$_9$ \cite{taniguchi},
(C$_5$H$_{12}$N$_2$)$_2$Cu$_2$Cl$_4$
\cite{chabou,collin1,collin2}, SrCu$_2$(BO$_3$)$_2$
\cite{ss1,ss2}, TlCuCl$_3$ \cite{tanaka,oosawa,ruegg,normand}, and
Cs$_2$CuCl$_4$ \cite{radu,radu2}. In some cases, it has even been
possible to tune these insulators across quantum phase transitions
by applied pressure \cite{oosawa} or by an applied magnetic field
\cite{collin1,collin2,tanaka,ruegg}. A useful survey of some of
these experiments may be found in the recent article by Matsumoto
{\em et al.} \cite{normand}.

It would clearly be valuable to understand the structure of the
global phase diagram of antiferromagnetic Mott insulators above
one dimension. The compounds mentioned above would then correspond
to distinct points in this phase diagram, and placing them in this
manner should help us better understand the relationship between
different materials. One could also classify the quantum critical
points accessed by the pressure or field-tuning experiments. The
purpose of this article is to review recent theoretical work
towards achieving this goal. We will focus mainly on the case of
two spatial dimensions ($d$), but our methods and results often
have simple generalizations to $d=3$.

One useful vantage point for opening this discussion is the family
of Mott insulators with a gap to all spin excitations. All spin
gap compounds discovered to date have the important property of
being ``dimerized'', or more precisely, they have an even number
of $S=1/2$ spins per unit cell \cite{slater}. In such cases, the
spin gap can be understood by adiabatic continuation from the
simple limiting case in which the spins form local spin singlets
within each unit cell. A simple approach that can be used for a
theoretical description of such insulators is the method of bond
operators \cite{bondops,chubukov}. This method has been widely
applied, and in some cases provides an accurate quantitative
description of numerical studies and experiments
\cite{kotov1,normand}. We will describe it here in
Section~\ref{sec:dimer} in the very simple context of a coupled
dimer antiferromagnet; similar results are obtained in more
complicated, and realistic, lattice structures.
Section~\ref{sec:dimer} will also describe the quantum phase
transition(s) accessed by varying coupling constants in the
Hamiltonian while maintaining spin rotation invariance (this
corresponds to experiments in applied pressure): the spin gap
closes at a quantum critical point beyond which there is magnetic
order. Section~\ref{sec:qc} will discuss some of the important
experimental consequences of this quantum criticality at finite
temperatures. A distinct quantum critical point, belonging to a
different universality class, is obtained when the spin gap is
closed by an applied magnetic field---this is described in
Section~\ref{sec:mag}.

The remaining sections discuss the theoretically much more
interesting and subtle cases of materials with an odd number of
$S=1/2$ spins per unit cell, such as La$_2$CuO$_4$ and
Cs$_2$CuCl$_4$. A complementary, but compatible, perspective on
the physics of such antiferromagnets may be found in the review
article by Misguich and Lhuillier \cite{misguich}.
Antiferromagnets in this class can develop a spin gap by {\em
spontaneously\/} breaking the lattice symmetry so that the lattice
is effectively dimerized (see discussion in the following
paragraph). There are no known materials with a spin gap in which
the lattice symmetry has not been broken, but there is a
theoretical consensus that spin gap states without lattice
symmetry breaking are indeed possible in $d>1$ \cite{lsm}. The
study of antiferromagnets with an odd number of $S=1/2$ spins per
unit cell is also important for the physics of the doped cuprates.
These materials exhibit spin-gap-like behavior at low dopings, and
many theories associate aspects of its physics with the spin gap
state proximate to the magnetically ordered state of the square
lattice antiferromagnet found in La$_2$CuO$_4$.

Section~\ref{sec:square} will describe the nature of a spin gap
state on the square lattice. We begin with the nearest-neighbor
$S=1/2$ Heisenberg Hamiltonian on the square lattice---this is
known to have a magnetic N\'{e}el order which breaks spin rotation
invariance. Now add further neighbor exchange couplings until
magnetic order is lost and a spin gap appears. We will show that
the ground state undergoes a novel, second-order quantum phase
transition to a state with {\em bond order}: translational
symmetry is spontaneously broken \cite{rs,rsb} so that the
resulting lattice structure has an even number of $S=1/2$ spins
per unit cell. So aspects of the non-zero spin excitations in this
paramagnet are very similar to the ``dimerized'' systems
considered in Section~\ref{sec:dimer}, and experimentally they
will appear to be almost identical. Indeed, it may well be that
the experimental materials initially placed in the class of
Section~\ref{sec:dimer}, are secretely systems in the class of
Section~\ref{sec:square} which have developed bond order driven by
the physics of antiferromagets (as in Section~\ref{sec:para}) at
some intermediate energy scale. The host lattice then distorts
sympathetically to the bond order, and is effectively dimerized.
Such materials will possess many more low-lying singlet
excitations than those in the theory of Section~\ref{sec:dimer}:
these excitations play an important role in the restoration of
translational symmetry as we move towards the N\'{e}el state.
Unfortunately, such singlet excitations are rather difficult to
detect experimentally.

Section~\ref{sec:triangle} will address the same issue as
Section~\ref{sec:square}, but for the case of the triangular
lattice. Here the spins are ordered in a non-collinear
configuration in the magnetically ordered state, as is observed at
low temperatures in Cs$_2$CuCl$_4$ \cite{radu,radu2}. We will
argue that in this case there is a route to destruction of
magnetic order in which the resulting spin gap state preserves
full lattice symmetry \cite{ReSaSpN,ijmp}. Such a spin gap state
has a novel `topological' order \cite{Wen} which endows its
excitations with charges under an emergent gauge force. Recent
experimental measurements of the dynamic structure factor of
Cs$_2$CuCl$_4$ appear to be described rather well by the
excitations of this topologically ordered state at energies above
which the magnetic order of the ground state emerges
\cite{radu2,ybkim}.

\section{Coupled dimer antiferromagnet}
\label{sec:dimer}

We begin by describing the quantum phase transition in a simple
two-dimensional model of antiferromagnetically coupled $S=1/2$
Heisenberg spins which has 2 spins per unit cell. The transition
is tuned by varying a dimensionless parameter $\lambda$. As we
noted in Section~\ref{sec:intro} different `dimerized' Mott
insulators will correspond to different values of $\lambda$, and
the value of $\lambda$ can be tuned by applying pressure
\cite{oosawa,normand}.

\begin{figure}[t]
\centering
\includegraphics[width=2.5in]{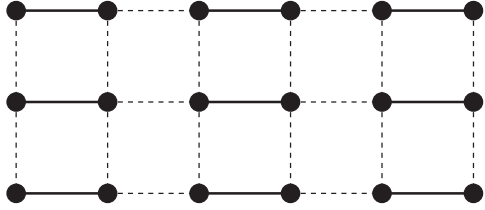}
\caption{ The coupled dimer antiferromagnet. Spins ($S=1/2$) are
placed on the sites, the $\mathcal{A}$ links are shown as full
lines, and the $\mathcal{B}$ links as dashed lines.} \label{fig1}
\end{figure}

We consider the ``coupled dimer'' Hamiltonian \cite{gsh}
\begin{equation}
H_{d} = J \sum_{\langle ij\rangle \in \mathcal{A}} {\bf S}_i \cdot
{\bf S}_j + \lambda J \sum_{\langle ij \rangle \in \mathcal{B}}
{\bf S}_i \cdot {\bf S}_j \label{ham}
\end{equation}
where ${\bf S}_j$ are spin-1/2 operators on the sites of the
coupled-ladder lattice shown in Fig~\ref{fig1}, with the
$\mathcal{A}$ links forming decoupled dimers while the
$\mathcal{B}$ links couple the dimers as shown.
The ground state of $H_d$ depends only on the dimensionless
coupling $\lambda$, and we will describe the low temperature ($T$)
properties as a function of $\lambda$. We will restrict our
attention to $J>0$ and $0 \leq \lambda \leq 1$.

Note that exactly at $\lambda =1$, $H_d$ is identical to the
square lattice antiferromagnet, and this is the only point at
which the Hamiltonian has only one spin per unit cell. At all
other values of $\lambda$ $H_d$ has a pair of $S=1/2$ spins in
each unit cell of the lattice. As will become clear from our
discussion, this is a key characteristic which permits a simple
theory for the quantum phase transition exhibited by $H_d$. Models
with only a single $S=1/2$ spin per unit cell usually display far
more complicated behavior, and will be discussed in
Sections~\ref{sec:square},\ref{sec:triangle}.

We will begin with a physical discussion of the phases and
excitations of the coupled dimer antiferromagnet, $H_{d}$ in
Section~\ref{sec:lad1}. We will propose a quantum
field-theoretical description of this model in
Section~\ref{sec:lad2}: we will verify that the limiting regimes
of the field theory contain excitations whose quantum numbers are
in accord with the phases discussed in Section~\ref{sec:lad1}, and
will then use the field theory to describe the quantum critical
behavior both at zero and finite temperatures.

\subsection{Phases and their excitations}
\label{sec:lad1}

Let us first consider the case where $\lambda$ is close to 1.
Exactly at $\lambda=1$, $H_d$ is identical to the square lattice
Heisenberg antiferromagnet, and this is known to have long-range,
magnetic N\'{e}el order in its ground state {\em i.e.} the
spin-rotation symmetry is broken and the spins have a non-zero,
staggered, expectation value in the ground state with
\begin{equation}
\langle {\bf S}_j \rangle = \eta_j N_0 {\bf n}, \label{neel}
\end{equation}
where ${\bf n}$ is some fixed unit vector in spin space, $\eta_j$
is $\pm 1$ on the two sublattices, and $N_0$ is the N\'{e}el order
parameter. This long-range order is expected to be preserved for a
finite range of $\lambda$ close to 1. The low-lying excitations
above the ground state consist of slow spatial deformations in the
orientation ${\bf n}$: these are the familiar spin waves, and they
can carry arbitrarily low energy {\em i.e.} the phase is
`gapless'. The spectrum of the spin waves can be obtained from a
text-book analysis of small fluctuations about the ordered
N\'{e}el state using the Holstein-Primakoff method
\cite{callaway}: such an analysis yields {\em two} polarizations
of spin waves at each wavevector $k = (k_x, k_y)$ (measured from
the antiferromagnetic ordering wavevector), and they have
excitation energy $\varepsilon_k = (c_x^2 k_x^2 + c_y^2
k_y^2)^{1/2}$, with $c_x, c_y$ the spin-wave velocities in the two
spatial directions.

\begin{figure}[t]
\centering
\includegraphics[width=2.5in]{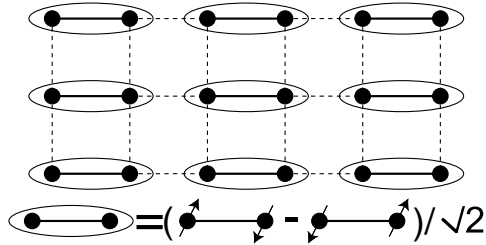}
\caption{Schematic of the quantum paramagnet ground state for
small $\lambda$. The ovals represent singlet valence bond pairs.
}\label{fig2}
\end{figure}

\begin{figure}[t]
\centering
\includegraphics[width=4.5in]{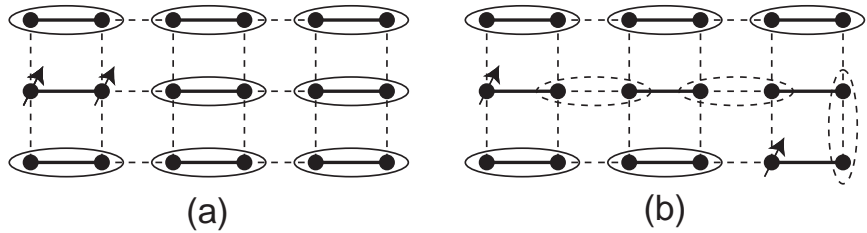}
\caption{(a) Cartoon picture of the bosonic $S=1$ excitation of
the paramagnet. (b) Fission of the $S=1$ excitation into two
$S=1/2$ spinons. The spinons are connected by a ``string'' of
valence bonds (denoted by dashed ovals) which lie on weaker bonds;
this string costs a finite energy per unit length and leads to the
confinement of spinons.} \label{fig3}
\end{figure}

Let us turn now to the vicinity of $\lambda = 0$. Exactly at
$\lambda=0$, $H_d$ is the Hamiltonian of a set of decoupled
dimers, with the simple exact ground state wavefunction shown in
Fig~\ref{fig2}: the spins in each dimer pair into valence bond
singlets, leading to a paramagnetic state which preserves spin
rotation invariance and all lattice symmetries.
Excitations are now formed by breaking a valence bond, which leads
to a {\em three}-fold degenerate state with total spin $S=1$, as
shown in Fig~\ref{fig3}a. At $\lambda=0$, this broken bond is
localized, but at finite $\lambda$ it can hop from site-to-site,
leading to a triplet quasiparticle excitation.
Note that this quasiparticle is {\em not\/} a spin-wave (or
equivalently, a `magnon') but is more properly referred to as a
spin 1 {\em exciton} or a {\em triplon} \cite{triplon}. We
parameterize its energy at small wavevectors $k$ (measured from
the minimum of the spectrum in the Brillouin zone) by
\begin{equation}
\varepsilon_k = \Delta + \frac{c_x^2 k_x^2 + c_y^2 k_y^2}{2
\Delta}, \label{epart}
\end{equation}
where $\Delta$ is the spin gap, and $c_x$, $c_y$ are velocities;
we will provide an explicit derivation of (\ref{epart}) in
Section~\ref{sec:lad2}. Fig~\ref{fig3} also presents a simple
argument which shows that the $S=1$ exciton cannot fission into
two $S=1/2$ `spinons'.

\begin{figure}[t]
\centering
\includegraphics[width=4.5in]{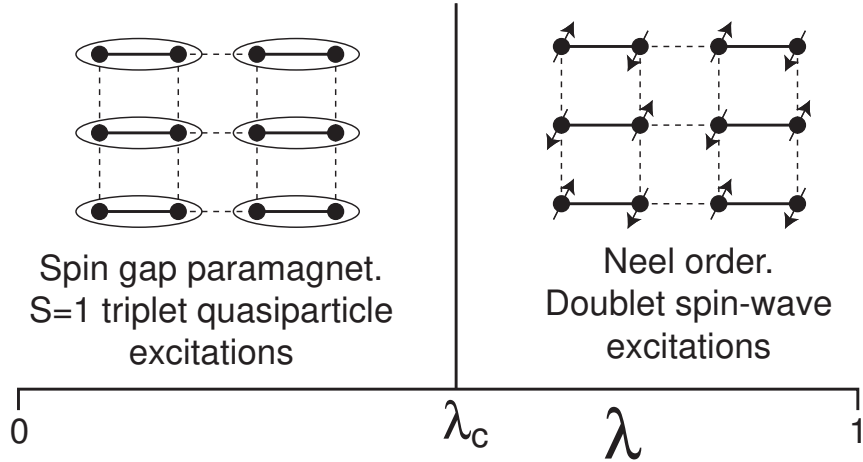}
\caption{Ground states of $H_d$ as a function of $\lambda$ The
quantum critical point is at \protect\cite{matsumoto} $\lambda_c =
0.52337(3)$. The compound TlCuCl$_3$ undergoes a similar quantum
phase transition under applied pressure \protect\cite{oosawa}.
}\label{fig4}
\end{figure}

The very distinct symmetry signatures of the ground states and
excitations between $\lambda \approx 1$ and $\lambda \approx 0$
make it clear that the two limits cannot be continuously
connected. It is known that there is an intermediate second-order
phase transition at \cite{gsh,matsumoto} $\lambda = \lambda_c =
0.52337(3)$ between these states as shown in Fig~\ref{fig4}.
Both the spin gap $\Delta$ and the N\'{e}el order parameter $N_0$
vanish continuously as $\lambda_c$ is approached from either side.

\subsection{Bond operators and quantum field theory}
\label{sec:lad2}

In this section we will develop a continuum description of the low
energy excitations in the vicinity of the critical point
postulated above. There are a number of ways to obtain the same
final theory: here we will use the method of {\em bond operators}
\cite{bondops,chubukov}, which has the advantage of making the
connection to the lattice degrees of freedom most direct. We
rewrite the Hamiltonian using bosonic operators which reside on
the centers of the $\mathcal{A}$ links so that it is explicitly
diagonal at $\lambda=0$. There are 4 states on each $\mathcal{A}$
link ($\left| \uparrow \uparrow \right\rangle$, $\left| \uparrow
\downarrow \right\rangle$, $\left| \downarrow \uparrow
\right\rangle$, and $\left| \downarrow \downarrow \right\rangle$)
and we associate these with the canonical singlet boson $s$ and
the canonical triplet bosons $t_{\alpha}$ ($\alpha=x,y,z$) so that
\begin{eqnarray}
|s \rangle \equiv  s^{\dagger} | 0 \rangle  =  \frac{1}{\sqrt{2}}
\left( |\uparrow\downarrow\rangle -
|\downarrow\uparrow\rangle\right) ~~;~~ |t_x \rangle \equiv
t_{x}^{\dagger} | 0 \rangle & = & \frac{-1}{\sqrt{2}} \left(
|\uparrow\uparrow\rangle - |\downarrow\downarrow\rangle\right)~~;
\nonumber \\
|t_y \rangle \equiv t_{y}^{\dagger} | 0 \rangle  =
\frac{i}{\sqrt{2}} \left( |\uparrow\uparrow\rangle +
|\downarrow\downarrow\rangle\right) ~~;~~ | t_z \rangle \equiv
t_{z}^{\dagger} | 0 \rangle & = & \frac{1}{\sqrt{2}} \left(
|\uparrow\downarrow\rangle + |\downarrow\uparrow\rangle\right).
\label{states}
\end{eqnarray}
Here $|0 \rangle$ is some reference vacuum state which does not
correspond to a physical state of the spin system. The physical
states always have a single bond boson and so satisfy the
constraint
\begin{equation}
s^{\dagger} s + t^{\dagger}_{\alpha} t_{\alpha} = 1 \label{const}
\end{equation}
By considering the various matrix elements $\langle s | {\bf S}_1
| t_{\alpha} \rangle$, $\langle s | {\bf S}_2 | t_{\alpha}
\rangle$, $\ldots$, of the spin operators ${\bf S}_{1,2}$ on the
ends of the link, it follows that the action of ${\bf S}_1$ and
${\bf S}_2$ on the singlet and triplet states is equivalent to the
operator identities
\begin{eqnarray}
S_{1\alpha} & = & \frac{1}{2}\left( s^{\dagger} t_{\alpha} +
t_{\alpha}^{\dagger} s - i\epsilon_{\alpha\beta\gamma}
t_{\beta}^{\dagger} t_{\gamma} \right), \nonumber  \\
S_{2\alpha} & = & \frac{1}{2}\left( - s^{\dagger} t_{\alpha} -
t_{\alpha}^{\dagger} s - i\epsilon_{\alpha\beta\gamma}
t_{\beta}^{\dagger} t_{\gamma} \right). \label{rep2}
\end{eqnarray}
where $\alpha$,$\beta$,$\gamma$ take the values $x$,$y$,$z$,
repeated indices are summed over and $\epsilon$ is the totally
antisymmetric tensor. Inserting (\ref{rep2}) into (\ref{ham}), and
using (\ref{const}), we find the following Hamiltonian for the
bond bosons:
\begin{eqnarray}
H_{d} &=& H_0 + H_1 \nonumber \\
H_0 &=& J \sum_{\ell \in \mathcal{A}} \left( - \frac{3}{4}
s_{\ell}^{\dagger} s_{\ell} + \frac{1}{4} t_{\ell\alpha}^{\dagger}
t_{\ell\alpha}
\right) \nonumber \\
H_1 &=& \lambda J \sum_{\ell,m \in \mathcal{A}} \Biggl[ a(\ell,m)
\left( t_{\ell \alpha}^{\dagger}  t_{m \alpha} s_{m}^{\dagger}
s_{\ell} + t_{\ell \alpha}^{\dagger}  t_{m \alpha}^{\dagger} s_{m}
s_{\ell} +
{\rm H.c.} \right) + b(\ell,m ) \nonumber \\
&~& \!\!\!\!\!\!\!\!\!\!\!\!\!\!\!\!\!\!\!\!\!\times \left( i
\epsilon_{\alpha\beta\gamma} t_{m \alpha}^{\dagger} t_{\ell
\beta}^{\dagger} t_{\ell \gamma} s_{m} + {\rm H.c.} \right) +
c(\ell,m ) \left( t_{\ell \alpha}^{\dagger} t_{m\alpha}^{\dagger}
t_{m\beta} t_{\ell\beta} - t_{\ell \alpha}^{\dagger}  t_{m
\beta}^{\dagger} t_{m\alpha} t_{\ell \beta} \right) \Biggr]
\label{bondham}
\end{eqnarray}
where $\ell,m$ label links in $\mathcal{A}$, and $a,b,c$ are
numbers associated with the lattice couplings which we will not
write out explicitly. Note that $H_1 =0 $ at $\lambda=0$, and so
the spectrum of the paramagnetic state is fully and exactly
determined. The main advantage of the present approach is that
application of the standard methods of many body theory to
(\ref{bondham}), while imposing the constraint (\ref{const}),
gives a very satisfactory description of the phases with $\lambda
\neq 0$, including across the transition to the N\'{e}el state. In
particular, an important feature of the bond operator approach is
that the simplest mean field theory already yields ground states
and excitations with the correct quantum numbers; so a strong
fluctuation analysis is not needed to capture the proper physics.

A complete numerical analysis of the properties of (\ref{bondham})
in a self-consistent Hartree-Fock treatment of the four boson
terms in $H_1$ has been presented in Ref.~\cite{bondops}. In all
phases the $s$ boson is well condensed at zero momentum, and the
important physics can be easily understood by examining the
structure of the low energy action for the $t_{\alpha}$ bosons.
For the particular Hamiltonian (\ref{ham}), the spectrum of the
$t_{\alpha}$ bosons has a minimum at the momentum $(0,\pi)$, and
for large enough $\lambda$ the $t_{\alpha}$ condense at this
wavevector: the representation (\ref{rep2}) shows that this
condensed state is the expected N\'{e}el state, with the magnetic
moment oscillating as in (\ref{neel}). The condensation transition
of the $t_{\alpha}$ is therefore the quantum phase transition
between the paramagnetic and N\'{e}el phases of the coupled dimer
antiferromagnet. In the vicinity of this critical point, we can
expand the $t_{\alpha}$ bose field in gradients away from the
$(0,\pi)$ wavevector: so we parameterize
\begin{equation}
t_{\ell, \alpha} (\tau) = t_{\alpha} (r_{\ell}, \tau) e^{i (0,\pi)
\cdot \vec r_{\ell}} \label{t1}
\end{equation}
where $\tau$ is imaginary time, $r\equiv (x,y)$ is a continuum
spatial co-ordinate, and expand the effective action in spatial
gradients. In this manner we obtain
\begin{eqnarray}
\mathcal{S}_t &=& \int d^2 r d \tau \left[ t^{\dagger}_{\alpha}
\frac{\partial t_{\alpha}}{\partial \tau} + C t^{\dagger}_{\alpha}
t_{\alpha} - \frac{D}{2} \left( t_{\alpha} t_{\alpha} + {\rm H.c.}
\right) + K_{1x} |\partial_x t_{\alpha} |^2 +
K_{1y} |\partial_y t_{\alpha} |^2\right. \nonumber \\
&~&~~~~~~~~~~~~~~~~~~~~\left. + \frac{1}{2} \left( K_{2x}
(\partial_x t_{\alpha} )^2 + K_{2y} (\partial_y t_{\alpha} )^2 +
{\rm H.c.} \right) + \cdots \right]. \label{st}
\end{eqnarray}
Here $C,D,K_{1,2 x,y}$ are constants that are determined by the
solution of the self-consistent equations, and the ellipses
represent terms quartic in the $t_{\alpha}$. The action
$\mathcal{S}_t$ can be easily diagonalized, and we obtain a $S=1$
quasiparticle excitation with the spectrum
\begin{equation}
\varepsilon_k = \left[ \left(C + K_{1x} k_x^2 + K_{1y} k_y^2
\right)^2 - \left( D + K_{2x} k_x^2 + K_{2y} k_y^2 \right)^2
\right]^{1/2}. \label{epart2}
\end{equation}
This is, of course, the triplon (or spin exciton) excitation of
the paramagnetic phase postulated earlier in (\ref{epart}); the
latter result is obtained by expanding (\ref{epart2}) in momenta,
with $\Delta = \sqrt{C^2 - D^2}$. This value of $\Delta$ shows
that the ground state is paramagnetic as long as $C>D$, and the
quantum critical point to the N\'{e}el state is at $C=D$.

The critical point and the N\'{e}el state are more conveniently
described by an alternative formulation of $\mathcal{S}_t$
(although an analysis using bond operators directly is also
possible \cite{sommer}). It is useful to decompose the complex
field $t_{\alpha}$ into its real and imaginary parts as follows
\begin{equation}
t_{\alpha} = Z (\varphi_{\alpha} + i \pi_{\alpha} ), \label{tphi}
\end{equation}
where $Z$ is a normalization chosen below. Insertion of
(\ref{tphi}) into (\ref{st}) shows that the field $\pi_{\alpha}$
has a quadratic term $\sim (C+D) \pi_{\alpha}^2$, and so the
co-efficient of $\pi_{\alpha}^2$ remains large even as the spin
gap $\Delta$ becomes small. Consequently, we can safely integrate
$\pi_{\alpha}$ out, and the resulting action for
$\varphi_{\alpha}$ takes the form
\begin{equation}
\mathcal{S}_{\varphi} = \int d^2 r d \tau \left[ \frac{1}{2}
\left\{ \left(
\partial_{\tau} \varphi_{\alpha} \right)^2 + c_x^2 \left(
\partial_{x} \varphi_{\alpha} \right)^2 + c_y^2 \left(
\partial_{y} \varphi_{\alpha} \right)^2 + s \varphi_{\alpha}^2
\right\} + \frac{u}{24} \left( \varphi_{\alpha}^2 \right)^2
\right]. \label{sp}
\end{equation}
Here we have chosen $Z$ to fix the co-efficient of the temporal
gradient term, and $s = C^2 - D^2$.

The formulation $\mathcal{S}_{\varphi}$ makes it simple to explore
the physics in the region $s<0$. It is clear that the effective
potential of $\varphi_{\alpha}$ has a minimum at a non-zero
$\varphi_{\alpha}$, and that $\langle \varphi_{\alpha} \rangle
\propto N_0$, the N\'{e}el order parameter in (\ref{neel}). It is
simple to carry out a small fluctuation analysis about this saddle
point, and we obtain the doublet of gapless spin-wave modes
advertised earlier.

We close this subsection by noting that all of the above results
have a direct generalization to other lattices, and also to spin
systems in three dimensions. Matsumoto {\em et al.} \cite{normand}
have applied the bond operator method to TlCuCl$_3$ and obtained
good agreement with experimental observations. One important
difference that emerges in such calculations on some frustrated
lattices \cite{balents} is worth noting explicitly here: the
minimum of the $t_{\alpha}$ spectrum need not be at special
wavevector like $(0,\pi)$, but can be at a more generic wavevector
$\vec{Q}$ such that $\vec{Q}$ and $-\vec{Q}$ are not separated by
a reciprocal lattice vector. A simple example which we consider
here is an extension of (\ref{ham}) in which there are additional
exchange interactions along all diagonal bonds oriented
`north-east' (so that the lattice has the connectivity of a
triangular lattice). In such cases, the structure of the low
energy action is different, as is the nature of the magnetically
ordered state. The parameterization (\ref{t1}) must be replaced by
\begin{equation}
t_{\ell \alpha} (\tau) = t_{1 \alpha} (r_{\ell}, \tau) e^{i \vec Q
\cdot \vec r_{\ell}} + t_{2 \alpha} (r_{\ell}, \tau) e^{-i \vec Q
\cdot \vec r_{\ell}} \label{t2}
\end{equation}
where $t_{1,2\alpha}$ are independent complex fields. Proceeding
as above, we find that the low energy effective action (\ref{sp})
is replaced by
\begin{eqnarray}
\mathcal{S}_{\Phi} &=& \int d^2 r d \tau \biggl[ \left|
\partial_{\tau} \Phi_{\alpha} \right|^2 + c_x^2 \left|
\partial_{x} \Phi_{\alpha} \right|^2 + c_y^2 \left|
\partial_{y} \Phi_{\alpha} \right|^2 + s
\left|\Phi_{\alpha}\right|^2 \nonumber \\
&~&~~~~~~~~~~~~~~~~~~~~~~~~~~ +\frac{u}{2} \left(
\left|\Phi_{\alpha} \right|^2 \right)^2 + \frac{v}{2} \left|
\Phi_{\alpha}^2 \right|^2 \biggr]. \label{sp2}
\end{eqnarray}
where now $\Phi_{\alpha}$ is a {\em complex} field such that
$\langle \Phi_{\alpha} \rangle \sim \langle t_{1\alpha} \rangle
\sim \langle t^{\dagger}_{2\alpha} \rangle$. Notice that there is
now a second quartic term with co-efficient $v$. If $v> 0$,
configurations with $\Phi_{\alpha}^2 = 0$ are preferred: in such
configurations $\Phi_{\alpha} = n_{1\alpha} + i n_{2 \alpha}$,
where $n_{1,2\alpha}$ are two equal-length orthogonal vectors.
Then from (\ref{t2}) and (\ref{rep2}) it is easy to see that the
physical spins possess {\em spiral} order in the magnetically
ordered state in which $\Phi_{\alpha}$ is condensed. A spiral
state is illustrated in Fig~\ref{fig13}, and we will have more to
say about this state in Section~\ref{sec:triangle}. For the case
$v < 0$, the optimum configuration has $\Phi_{\alpha} = n_{\alpha}
e^{i \theta}$ where $n_{\alpha}$ is a real vector: this leads to a
magnetically ordered state with spins polarized {\em collinearly}
in a spin density wave at the wavevector $\vec{Q}$.

\subsection{Quantum criticality}
\label{sec:qc}

We will restrict our discussion here to the critical point
described by $\mathcal{S}_{\varphi}$. Similar results apply to
$\mathcal{S}_\Phi$ for the parameter regime in which it exhibits a
second order transition \cite{kawamura}. Experimentally, the
results below are relevant to materials that can be tuned across
the magnetic ordering transition by applied pressure (such as
TlCuCl$_3$ \cite{oosawa}), or to materials which happen to be near
a critical point at ambient pressure (such as LaCuO$_{2.5}$
\cite{bruce}).

The field theory $\mathcal{S}_{\varphi}$ is actually a familiar
and well-studied model in the context of classical critical
phenomena. Upon interpreting $\tau$ as a third spatial
co-ordinate, $\mathcal{S}_{\varphi}$ becomes the theory of a
classical O(3)-invariant Heisenberg ferromagnet at finite
temperatures (in general a $d$ dimensional quantum antiferromagnet
will map to a $d+1$ dimensional classical Heisenberg ferromagnet
at finite temperature \cite{chn}). The Curie transition of the
Heisenberg ferromagnet then maps onto the quantum critical point
between the paramagnetic and N\'{e}el states described above. A
number of important implications for the quantum problem can now
be drawn immediately.

The theory $\mathcal{S}_{\varphi}$ has a `relativistic'
invariance, and consequently the dynamic critical exponent must be
$z=1$. The spin correlation length will diverge at the quantum
critical point with the exponent \cite{landau} $\nu = 0.7048(30)$.
The spin gap of the paramagnet, $\Delta$, vanishes as $\Delta \sim
(\lambda_c - \lambda)^{z \nu}$, and this prediction is in
excellent agreement with the numerical study of the dimerized
antiferromagnet \cite{matsumoto}.

A somewhat more non-trivial consequence of this mapping is in the
structure of the spectrum at the critical point $\lambda =
\lambda_c$. At the Curie transition of the classical ferromagnet
it is known \cite{ma} that spin correlations decay as $\sim
1/p^{2-\eta}$, where $p$ is the 3-component momentum in the
3-dimensional classical space. We can now analytically continue
this expression from its $p_z$ dependence in the third classical
dimension to the real frequency, $\omega$, describing the quantum
antiferromagnet. This yields the following fundamental result for
the dynamic spin susceptibility, $\chi (k, \omega)$, at the $T=0$
quantum critical point of the coupled-dimer antiferromagnet:
\begin{equation}
\chi (k, \omega ) \sim \frac{1}{\left( c_x^2 k_x^2 + c_y^2 k_y^2 -
(\omega+ i\epsilon)^2 \right)^{1-\eta/2}}, \label{e4}
\end{equation}
where $\epsilon$ is a positive infinitesimal. Note that in
(\ref{e4}) the momentum $k$ is measured from the $(\pi,\pi)$
ordering wavevector of the N\'{e}el state. The exponent $\eta$ is
the same as that of the classical Heisenberg ferromagnet, and has
a rather small value \cite{landau}: $\eta \approx 0.03$. However,
the non-zero $\eta$ does make a significant difference to the
physical interpretation of the excitations at the critical point.
In particular note that $\mbox{Im} \chi (k, \omega)$ does not have
a pole at any $k$, but rather a continuum spectral weight above a
threshold energy \cite{sy,csy}
\begin{equation}
\mbox{Im} \chi (k, \omega) \sim \mbox{sgn}(\omega)
\sin\left(\frac{\pi\eta}{2} \right) \frac{\theta\left(|\omega| -
\sqrt{c_x^2 k_x^2 + c_y^2 k_y^2}\right)}{\left(\omega^2 - c_x^2
k_x^2 - c_y^2 k_y^2 \right)^{1-\eta/2}}
\end{equation}
where $\theta$ is the unit step function. This indicates there are
no quasiparticles at the critical point, and only a dissipative
critical continuum.

There is also some very interesting structure in the quantum
critical dynamic response at nonzero $T$ \cite{sy,csy}. Here, one
way to understand the physics is to approach the critical point
from the paramagnetic side ($\lambda < \lambda_c$). As we noted
earlier, the paramagnetic phase has well-defined `triplon' or
`spin exciton' excitations $t_{\alpha}$, and these have an
infinite lifetime at $T=0$. At $T>0$, thermally excited
$t_{\alpha}$ quasiparticles will collide with each other via their
scattering amplitude, $u$, and this will lead to a finite lifetime
\cite{csy,damle}. Now approach $\lambda = \lambda_c$. The
renormalization group analysis of $\mathcal{S}_{\varphi}$ tells us
that the quartic coupling $u$ approaches a fixed point value in
the critical region. This means that $u$ is no longer an arbitrary
parameter, and an appropriately defined $t_{\alpha}$ scattering
amplitude must also acquire universal behavior. In particular, the
$t_{\alpha}$ lifetime is determined by the only energy scale
available, which is $k_B T$. So we have the remarkable result that
the characteristic spin relaxation time is a universal number
times $\hbar/(k_B T)$. More precisely, we can write for the local
dynamic spin susceptibility $\chi_L (\omega) = \int d^2 k \chi(k,
\omega)$ the universal scaling form
\begin{equation}
\mbox{Im} \chi_L (\omega) = T^{\eta} F \left(\frac{\hbar
\omega}{k_B T} \right)
\end{equation}
Here $F$ is a universal function which has the limiting behaviors
\begin{equation}
F(\overline{\omega}) \sim \left\{ \begin{array}{lll}
\overline{\omega} &,& |\overline{\omega}| \ll 1 \\
\mbox{sgn} (\omega) |\overline{\omega}|^{\eta} &,&
|\overline{\omega}| \gg 1
\end{array} \right. .
\end{equation}
Note that $F$ has a smooth linear behavior in the regime $|\hbar
\omega| \ll k_B T$, and this is similar to any simple dissipative
system. The difference here is that the co-efficient of
dissipation is determined by $k_B T$ alone.

The quantum critical behavior described here is expected to apply
more generally to other correlated electron systems, provided the
critical theory has non-linear couplings which approach fixed
point values.

\section{Influence of an applied magnetic field}
\label{sec:mag}

An important perturbation that can be easily applied to
antiferromagnets in the class discussed in Section~\ref{sec:dimer}
is a uniform magnetic field. The Zeeman energy in available fields
can often be comparable to the typical antiferromagnetic exchange
constant $J$, and so the ground state can be perturbed
significantly. It is therefore of interest to understand the
evolution of the phase diagram in Fig~\ref{fig4} under an applied
field of arbitrary strength.

We are interested here in the evolution of the ground state as a
function of $B$ where the Hamiltonian $H_d$ in (\ref{ham}) is
transformed as
\begin{equation}
H_d \rightarrow H_d - \sum_j {\bf B} \cdot {\bf S}_{j}.
\label{hamb}
\end{equation}
Most of the basic features can actually be understood quite easily
in a simple extension of the self-consistent Hartree-Fock theory
of bond bosons that was discussed in Section~\ref{sec:lad2}. Under
the transformation (\ref{hamb}), it is easily seen from
(\ref{rep2}) that
\begin{equation}
H_d \rightarrow H_d + i B_{\alpha} \sum_{\ell \in \mathcal{A}}
\epsilon_{\alpha\beta\gamma} t_{\ell \beta}^{\dagger} t_{\ell
\gamma} \label{hambt}
\end{equation}
The presence of a non-zero $B$ breaks spin rotation invariance and
so all the self-consistent expectation values of operator
bilinears have to reflect this reduced symmetry in the
Hartree-Fock theory. Apart from this the mechanics of the
computation mostly remain the same. However, for stronger fields,
it is sometimes necessary to allow for broken translational
symmetry in the expectation values, as the ground state can
acquire a modulated structure.

We will discuss the results of such an analysis in weak and strong
fields in the following subsections.

\subsection{Weak fields}
\label{sec:weak}

\begin{figure}[t]
\centering
\includegraphics[width=4.5in]{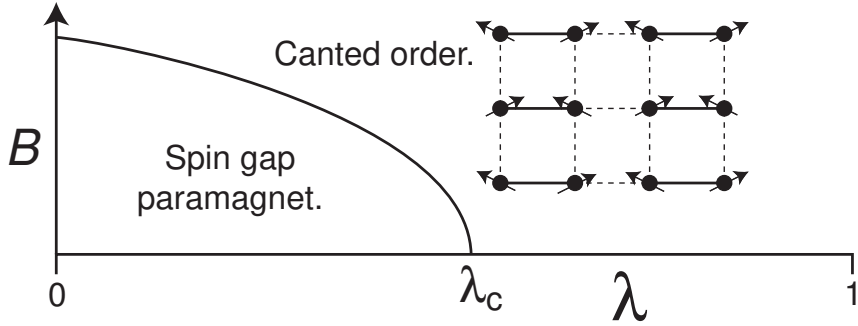}
\caption{Evolution of the phases of Fig~\protect\ref{fig4} under a
weak field $B$ (magnetization plateau at large $B$, appearing in
Fig~\protect\ref{fig6}, are not shown). The paramagnetic phase has
exactly the same ground state wavefunction as that at $B=0$. The
phase boundary behaves like $B \sim (\lambda_c - \lambda)^{z
\nu}$. The $B$ field is oriented vertically upwards, and the
static moments in the canted phase can rotate uniformly about the
vertical axis. The phase boundary at non-zero $B$ is described by
the $z=2$ dilute Bose gas quantum critical theory. The phase
diagram of TlCuCl$_3$ in applied pressure and magnetic field looks
similar to the one above \protect\cite{normand}. The corresponding
phase diagram of the field-induced magnetic ordering transition of
a superconductor (rather than a Mott insulator) has been
investigated recently \protect\cite{demler}, and successfully
applied to experiments on the doped cuprates; this phase diagram
of the superconductor has significant differences from the one
above.}\label{fig5}
\end{figure}

For weak fields applied to the paramagnet (specifically, for
fields $B < \Delta$, the influence of (\ref{hambt}) can be
understood exactly. The coupling to $B$ involves an operator which
commutes with the remaining Hamiltonian (the total spin), and
hence the wavefunction of the ground state remains insensitive to
the value of $B$. The same applies to the wavefunctions of the
excited states. However, the excited states can have non-zero
total spin and so their energies do depend upon $B$. In particular
the triplet $t_{\alpha}$ quasiparticle with energy (\ref{epart})
or (\ref{epart2}) carries total spin $S=1$, and consequently we
conclude that this triplet splits according to
\begin{equation}
\varepsilon_k \rightarrow \varepsilon_k - m B
\end{equation}
with $m=0,\pm 1$. Note that the lowest energy quasiparticle (with
$m=1$) has a positive energy as long as $B < \Delta$, and this is
required for the stability of the paramagnet. So the phase
boundary of the paramagnetic phase is exactly $B=\Delta$, and
using $\Delta \sim (\lambda_c - \lambda )^{z \nu}$, we can sketch
the boundary of the paramagnetic phase as in Fig~\ref{fig5}.

What happens beyond the paramagnetic phase? As in
Section~\ref{sec:lad2}, we answer this question by using the
transformation (\ref{tphi}), and by examining the analog of
$\mathcal{S}_{\varphi}$ under a non-zero $B$. Using (\ref{hambt}),
and integrating out $\pi_{\alpha}$, we now find that the action
$\mathcal{S}_{\varphi}$ in (\ref{sp}) remains unchanged apart from
the mapping \cite{affleck}
\begin{equation}
\left(
\partial_{\tau} \varphi_{\alpha} \right)^2 \rightarrow
\left(
\partial_{\tau} \varphi_{\alpha} + i \epsilon_{\alpha\beta\gamma} B_{\beta} \varphi_{\gamma}
\right)^2 . \label{spb}
\end{equation}
The action (\ref{sp}), (\ref{spb}) can now be analyzed by a
traditional small fluctuation analysis about $\varphi_{\alpha}=0$.
Let us assume that ${\bf B} = (0,0,B)$ is oriented along the $z$
axis. Then the co-efficient of $\varphi_z^2$ is $s$, while that of
$\varphi_x^2 + \varphi_y^2$ is $s-B^2$. This suggests that we
focus only on the components of $\varphi_{\alpha}$ in the plane
orthogonal to ${\bf B}$, and integrate out the component of
$\varphi_{\alpha}$ along the direction of ${\bf B}$. Indeed, if we
define
\begin{equation}
\Psi = \frac{\varphi_x + i \varphi_y}{\sqrt{B}} \label{psiphi}
\end{equation}
and integrate out $\varphi_z$, then we obtain from (\ref{sp}),
(\ref{spb}) the effective action for $\Psi$:
\begin{equation}
\mathcal{S}_{\Psi} = \int d^2 r d \tau \left[ \Psi^{\ast}
\partial_{\tau} \Psi  + \frac{c_x^2}{2B} \left|
\partial_{x} \Psi \right|^2 + \frac{c_y^2}{2B} \left|
\partial_{y} \Psi \right|^2 -\mu |\Psi|^2
 + \frac{u}{24B} |\Psi|^4 \right].
\label{spsi}
\end{equation}
Here, $\mu=(s-B^2)/2B$, and we have retained only leading order
temporal and spatial gradients and the leading dependence of $u$.
Clearly, this is the theory of a Bose gas in the grand canonical
ensemble at a chemical potential $\mu$, with a repulsive
short-range interaction \cite{sss}. At $T=0$, and $\mu < 0$, such
a theory has a ground state which is simply the vacuum with no
Bose particles. Here, this vacuum state corresponds to the spin
gap antiferromagnet, and the $B$-independence of the ground state
of the antiferromagnet corresponds here to the $\mu$ independence
of the ground state of $\mathcal{S}_{\Psi}$. There is an onset of
a finite density of bosons in $\mathcal{S}_{\Psi}$ for $\mu > 0$,
and this onset therefore corresponds to the quantum phase
transition in the antiferromagnet at $B=\Delta$. So we must have
$\mu=0$ in $\mathcal{S}_{\Psi}$ at precisely the point where
$B=\Delta$: the value of $\mu$ quoted above shows that this is
true at zeroth order in $u$, and higher order terms in $u$ must
conspire to maintain this result.

The above analysis makes it clear that the $\mu \geq 0$ region of
$\mathcal{S}_{\Psi}$ will describe the quantum phase transition
out of the paramagnet at non-zero $B$. This transition is merely
the formation of a Bose-Einstein condensate of the $m=1$ component
of the triplon bosons. For $\mu
> 0$ we have a finite density of $\Psi$ bosons which Bose condense
in the ground state, so that $\langle \Psi \rangle \neq 0$. From
(\ref{psiphi}) we see that this Bose condensation corresponds to
antiferromagnetic order in the plane perpendicular to ${\bf B}$.
Indeed, the phase of this Bose condensate is simply the
orientation of the spins in the $x,y$ plane, and so here this
phase is directly observable. Further, by taking derivatives of
(\ref{hamb}) and $\mathcal{S}_\Psi$ w.r.t. $B$, we see that the
density of bosons is proportional to the magnetization per spin,
$\Omega$, in the direction parallel to ${\bf B}$:
\begin{equation}
\Omega \equiv \frac{1}{N} \sum_j \langle S_{jz} \rangle \propto
\langle |\Psi|^2 \rangle, \label{Omega}
\end{equation}
where $N$ is the total number of spins. Consequently, the average
magnetic moments in the non-paramagnetic phase are in a `canted'
configuration, as shown in Fig~\ref{fig5}. The quantum phase
transition between the paramagnet and the canted state is
described by the theory of the density onset in a Bose gas: this
theory has $z=2$, $\nu=1/2$, and an upper critical dimension of
$d=2$ \cite{fwgf,sss}.

We conclude this section by noting that interesting recent work
\cite{shindo} has examined the Bose-Einstein condensation of the
$m=1$ triplon bosons in a {\em random} potential. This is achieved
by studying Tl$_{1-x}$K$_x$CuCl$_3$, where the stoichiometric
disorder among the non-magnetic ions acts as a random potential on
the triplons.

\subsection{Strong fields}
\label{sec:strong}

We have seen above that applying a magnetic field eventually leads
to the onset of a ferromagnetic moment in the directions of the
applied field. How does this moment evolve as we continue to
increase the field? Eventually, $B$ will become so large that it
pays to have all the spins polarized in the direction of the
field: this corresponds to a saturation in the magnetization, and
making $B$ even stronger will not change the ground state. In
terms of the $t$ bosons, this fully polarized state, $|FP
\rangle$, with $\Omega=1/2$, is seen from (\ref{hambt}) or
(\ref{states}) to correspond exactly to
\begin{equation}
|FP \rangle = \prod_{\ell} \frac{(t^{\dagger}_{\ell x} + i
t^{\dagger}_{\ell y})}{\sqrt{2}} |0\rangle.
\end{equation}
So there must be at least one more quantum phase transition as a
$B$ is increased: this is transition from the $|FP \rangle$ state
at very large $B$ to a state with a continuously varying
ferromagnetic moment which eventually reaches the saturation value
from below.

A theory for the transition away from the $|FP \rangle$ state with
decreasing $B$ can be developed using methods very similar to
those used in Section~\ref{sec:lad2} and~\ref{sec:weak}. We treat
the quartic terms in (\ref{bondham}) in a Hartree-Fock
approximation, and examine small fluctuations away from the $| FP
\rangle$ state. These are dominated by excitation which create
$t_z$ quanta (which have $m=0$) on the dimers, and so the
effective theory is expressed in terms of
\begin{equation}
\widetilde{\Psi}^{\dagger} \sim t^{\dagger}_z (t_x - i t_y).
\end{equation}
Indeed, it is not difficult to see that the resulting theory for
$\widetilde{\Psi}$ has exactly the same form as (\ref{spsi}). Now
the $\mu$ for $\widetilde{\Psi}$ decreases with increasing $B$,
and we have $\mu=0$ at the critical field at which $|FP \rangle$
first becomes the ground state. Furthermore, $\langle
|\widetilde{\Psi} |^2 \rangle$ now measures the deviation away
from $\Omega=1/2$. Apart from this `inversion' in the field axis,
it is clear that the universality class of the present transition
is identical to that discussed in Section~\ref{sec:weak}.

A further possibility for a plateau in the value of $\Omega$ with
increasing $B$ is worth mentioning \cite{oshikawa}, as analogs are
realized in SrCu$_2$(BO$_3$)$_2$ \cite{plateauss} and
NH$_4$CuCl$_3$ \cite{nh4}. So far we have found plateaus at
$\Omega =0$ for $B< \Delta$, and at $\Omega=1/2$ for large $B$.
For the $\Omega=1/2$ state we had every dimer with a
$(t^{\dagger}_x + i t^{\dagger}_y)/\sqrt{2}$ boson. Now imagine
that these bosons form a Wigner-crystalline state so that there
are $p$ such bosons for every $q$ dimers; here $0 \leq p \leq q$,
$q \geq 1$, are integers. Such a state will have $\Omega = p/(2
q)$, and breaks the translational symmetry of the underlying dimer
antiferromagnet such that there are $q$ dimers per unit cell (or
$2q$ spins per unit cell). The energy gap towards boson motion in
the Wigner crystal ({\em i.e.} its incompressibility) will ensure
that $\Omega$ is stable under small variations of $B$. In this
manner we can obtain a magnetization plateau at $\Omega=p/(2q)$ in
a state with a unit cell of $q$ dimers.

\begin{figure}[t]
\centering
\includegraphics[width=2.5in]{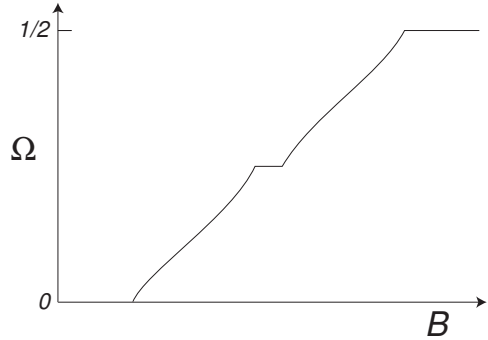}
\caption{Magnetization density, $\Omega$, defined in
(\protect\ref{Omega}) as a function of the applied magnetic field.
The plateau shown at $\Omega=0$ is present provided the zero field
state is a paramagnet {\em i.e.\/} $\lambda < \lambda_c$. The full
saturation plateau at $\Omega=1/2$ is always present. The plateau
at $\Omega=1/4$ is not present in the nearest-neighbor model $H_d$
in (\ref{ham}), but it is believed that such a plateau will appear
upon including frustrating exchange interactions; this plateau
will involve a broken translational symmetry in the coupled dimer
antiferromagnet. Such magnetization plateaux are found in
SrCu$_2$(BO$_3$)$_2$ \protect\cite{plateauss} and NH$_4$CuCl$_3$
\protect\cite{nh4} }\label{fig6}
\end{figure}

We summarize the considerations of this subsection in
Fig~\ref{fig6}, showing a possible evolution of $\Omega$ in a
model similar to $H_d$ in (\ref{ham}).
As we have already noted, the plateau onset transitions at $\Omega
=0$ and $\Omega=1/2$ are both described by the $z=2$ dilute Bose
gas theory (\ref{spsi}). The transitions in and out of other
fractional plateaus are potentially more complicated because these
involve spontaneous breaking of translational symmetry. The
translation symmetry could be restored at the same point at which
there is onset of superfluid order---this is possibly a first
order transition with a jump in the value of $\Omega$.
Alternatively, there could be an intermediate `supersolid' phase,
in which case the plateau transition has the same broken
translational symmetry on both sides of it, placing it also in the
class of (\ref{spsi}).

\section{Square lattice antiferromagnet}
\label{sec:square}

This section will address the far more delicate problem of quantum
phase transitions in antiferromagnets with an odd number of
$S=1/2$ spins per unit cell. We will mainly concern ourselves with
square lattice Hamiltonians of the form
\begin{equation}
H_s = J\sum_{\langle ij\rangle} {\bf S}_i \cdot {\bf S}_j + \ldots
. \label{hams}
\end{equation}
Here $J$ is a nearest-neighbor antiferromagnetic exchange and the
ellipses represent further short-range exchange interactions
(possibly involving multiple spin ring exchange) which preserve
the full symmetry of the square lattice. The model $H_d$ is a
member of the class $H_s$ only at $\lambda=1$; at other values of
$\lambda$ the symmetry group of the square lattice is explicitly
broken, and the doubling of the unit cell was crucial in the
analysis of Section~\ref{sec:dimer}. With full square lattice
symmetry, the paramagnetic phase is not determined as simply as in
the small $\lambda$ expansion, and we have to account more
carefully for the `resonance' between different valence bond
configurations.

One ground state of $H_s$ is, of course, the N\'{e}el state
characterized by (\ref{neel}); this is obtained in the absence of
the interactions denoted by ellipses in (\ref{hams}). Now imagine
tuning the further neighbor couplings in (\ref{hams}) so that spin
rotation invariance is eventually restored and we obtain a
paramagnetic ground state. We can divide the possibilities for
this state into two broad classes, which we discuss in turn.

In the first class of paramagnets, no symmetries of the
Hamiltonian are broken, and the spins have paired with each other
into valence bond singlets which strongly resonate between the
large number of possible pairings: this is a resonating valence
bond (RVB) liquid \cite{pwa,krs}. We will discuss such states
further in Section~\ref{sec:triangle}: they have a connection with
magnetically ordered states with non-collinear magnetic order,
unlike the collinear N\'{e}el state of the nearest neighbor square
lattice antiferromagnet.

In the second class of paramagnets, the valence bond singlets
spontaneously crystallize into some configuration which
necessarily breaks a lattice symmetry. A simple example of such a
{\em bond-ordered} paramagnet is the columnar state we have
already considered in Fig~\ref{fig2}. For the dimerized
antiferromagnet $H_d$, the bond configuration in Fig~\ref{fig2}
was chosen explicitly in the Hamiltonian by the manner in which we
divided the links into classes $\mathcal{A}$ and $\mathcal{B}$ for
$\lambda \neq 1$. For $H_s$, there is no such distinction between
the links, and hence a state like Fig~\ref{fig2} {\em
spontaneously} breaks a lattice symmetry. Furthermore, there are 3
other equivalent states, obtained by successive 90 degree
rotations of Fig~\ref{fig2} about any lattice site, which are
completely equivalent. So for $H_s$, the bond-ordered paramagnet
in Fig~\ref{fig2} is four-fold degenerate. Going beyond simple
variational wavefunctions like Fig~\ref{fig2}, the bond-ordered
states are characterized by a bond order parameter
\begin{equation}
Q_{ij} = \langle {\bf S}_i \cdot {\bf S}_j \rangle;
\label{bondorder}
\end{equation}
the values of $Q_{ij}$ on the links of the lattice in a
bond-ordered state have a lower symmetry than the values of the
exchange constants $J_{ij}$ in the Hamiltonian. We will develop an
effective model for quantum fluctuations about the collinear
N\'{e}el state in $H_s$ below, and will find that such
bond-ordered paramagnets emerge naturally \cite{rs}.

Let us now try to set up a theory for quantum fluctuations about
the N\'{e}el state (\ref{neel}). It is best to do this in a
formulation that preserves spin rotation invariance at all stages,
and this is facilitated by the coherent state path integral (see
Chapter 13 of Ref.~\cite{book}). The essential structure of this
path integral can be understood simply by looking at a single spin
in a magnetic field ${\bf h}$ with the Hamiltonian $H_1 = - {\bf
h} \cdot {\bf S}$. Then its partition function at a temperature
$T$ is given by
\begin{equation}
\mbox{Tr} \exp \left({\bf h} \cdot {\bf S}/T \right) = \int
\mathcal{D} {\bf n} (\tau) \exp \left( i 2S \mathcal{A} [ {\bf n}
(\tau)] + S\int_0^{1/T} d \tau {\bf h} \cdot {\bf n} (\tau)
\right). \label{single}
\end{equation}
Here $S$ is the angular momentum of the spin ${\bf S}$ (we are
interested primarily in the case $S=1/2$) and ${\bf n} (\tau)$ is
a unit 3-vector with ${\bf n} (0) = {\bf n} (1/T)$. So the above
path integral is over all closed curves on the surface of a
sphere. The first term in the action of the path integral is the
crucial Berry phase: $\mathcal{A}[{\bf n}(\tau)]$ is {\em half}
the oriented area enclosed by the curve ${\bf n} (\tau )$ (the
reason for the half will become clear momentarily). Note that this
area is only defined modulo $4 \pi$, the surface area of a unit
sphere. The expression (\ref{single}) has an obvious
generalization to the lattice Hamiltonian $H_s$: the action adds
up the Berry phases of every spin, and there is an additional
energy term which is just the Hamiltonian with the replacement
${\bf S}_j \rightarrow S {\bf n}_j$.

\begin{figure}[t]
\centering
\includegraphics[width=2.5in]{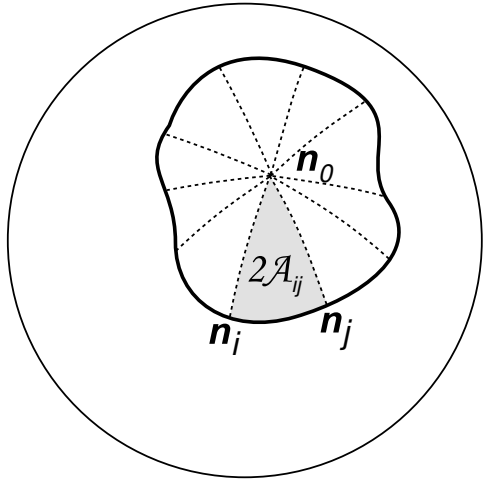}
\caption{The path traced out by a single spin on the unit sphere
in imaginary time. After discretizing time, the area enclose by
the path is written as the sum over the areas of spherical
triangles: $\mathcal{A}_{ij}$ is half the area of the triangle
with vertices ${\bf n}_0$, ${\bf n}_i$, ${\bf n}_j$. Different
choices for the arbitrary point ${\bf n}_0$ correspond to
different gauge choices associated with (\protect\ref{gauge1}) and
(\protect\ref{gauge2}).}\label{fig7}
\end{figure}

We are now faced with the problem of keeping track of the areas
enclosed by the curves traced out by all the spins. This seems
rather daunting, particularly because the half-area $\mathcal{A}[
{\bf n} (\tau)]$ is a global object defined by the whole curve,
and cannot be obviously be associated with local portions of the
curve. One convenient way to proceed is illustrated in
Fig~\ref{fig7}: discretize imaginary time, choose a fixed
arbitrary point ${\bf n}_0$ on the sphere, and thus write the area
as the sum of a large number of spherical triangles. Note that
each triangle is associated with a local portion of the curve
${\bf n}(\tau)$.

We now need an expression for $\mathcal{A} ( {\bf n}_1, {\bf n}_2
, {\bf n}_3 )$, defined as half the area of the spherical triangle
with vertices ${\bf n}_1$, ${\bf n}_2$, ${\bf n}_3$. Complicated
expressions for this appear in treatises on spherical
trigonometry, but a far simpler expression is obtained after
transforming to spinor variables \cite{berg}. Let us write
\begin{equation}
{\bf n}_j \equiv z^{\ast}_{ja} \vec{\sigma}_{ab} z_{jb} \label{nz}
\end{equation}
where $a,b = \uparrow, \downarrow$ and we will always assume an
implied summation over such indices, $\vec{\sigma}_{ab}$ are the
Pauli matrices, and $z_{j\uparrow}$, $z_{j\downarrow}$ are complex
numbers obeying $|z_{j\uparrow}|^2 + |z_{j\downarrow}|^2 = 1$.
Note that knowledge of ${\bf n}_j$ only defines $z_{ja}$ up to a
U(1) gauge transformation under which
\begin{equation}
z_{ja} \rightarrow z_{ja} e^{i \phi_j}. \label{gauge1}
\end{equation}
Then, associated with each pair of vertices ${\bf n}_{i},{\bf
n}_j$ we define
\begin{equation}
\mathcal{A}_{ij} \equiv \arg \left[ z_{i a}^{\ast} z_{j a}
\right]. \label{Az}
\end{equation}
Under the gauge transformation (\ref{gauge1}) we have
\begin{equation}
\mathcal{A}_{ij} \rightarrow \mathcal{A}_{ij} - \phi_i + \phi_j,
\label{gauge2}
\end{equation}
{\em i.e.\/} $\mathcal{A}_{ij}$ behaves like a U(1) gauge field.
Note also that $\mathcal{A}_{ij}$ is only defined modulo $2 \pi$,
and that $\mathcal{A}_{ji} = -\mathcal{A}_{ij}$. For future use,
we also mention the following identity, which follows from
(\ref{nz}) and (\ref{Az}):
\begin{equation}
z_{i a}^{\ast} z_{j a} = \left( \frac{1 + {\bf n}_i \cdot {\bf
n}_j }{2} \right)^{1/2} e^{i \mathcal{A}_{ij}}. \label{zzA}
\end{equation}
The classical result for the half-area of the spherical triangle
can be written in the simple form in terms of the present U(1)
gauge variables:
\begin{equation}
\mathcal{A} ({\bf n}_1 , {\bf n}_2 , {\bf n}_3 ) =
\mathcal{A}_{12} + \mathcal{A}_{23} + \mathcal{A}_{31}
\label{triangle}
\end{equation}
We chose $\mathcal{A}$ as a {\em half}-area earlier mainly because
then the expressions (\ref{Az}) and (\ref{triangle}) come out
without numerical factors. It is satisfying to observe that this
total area is invariant under (\ref{gauge2}), and that the
half-area is ambiguous modulo $2 \pi$.

Using (\ref{triangle}), we can now write down a useful expression
for $\mathcal{A}[{\bf n} (\tau)]$. We assume that imaginary time
is discretized into times $\tau_j$ separated by intervals $\Delta
\tau$. Also, we denote by $j+\tau$ the site at time $\tau_j +
\Delta \tau$, and define $\mathcal{A}_{j,j+\tau} \equiv
\mathcal{A}_{j\tau}$. Then
\begin{equation}
\mathcal{A} [ {\bf n} (\tau)] = \sum_j \mathcal{A}_{j \tau}
\label{e1}
\end{equation}
Note that this expression is a gauge-invariant function of the
U(1) gauge field $\mathcal{A}_{j \tau}$, and is analogous to the
quantity sometimes called the Polyakov loop.

We are now ready to write down the first form proposed effective
action for the quantum fluctuating N\'{e}el state. We do need to
address some simple book-keeping considerations first:\\
({\em i\/}) Discretize spacetime into a cubic lattice of points
$j$. Note that the same index $j$ referred to points along
imaginary time above, and to square lattice points in $H_s$. The
meaning of
the site index should be clear from the context. \\
({\em ii\/}) On each spacetime point $j$, we represent quantum
spin operator ${\bf S}_j$ by
\begin{equation}
{\bf S}_j = \eta_j S {\bf n}_j ,\label{Sn}
\end{equation}
where ${\bf n}_j$ is a unit vector, and $\eta_j=\pm 1$ is the
sublattice staggering factor appearing in (\ref{neel}). This
representation is that expected from the coherent state path
integral, apart from the $\eta_j$ factor. We have chosen to
include $\eta_j$ because of the expected local antiferromagnetic
correlations of the spins. So in a quantum fluctuating N\'{e}el
state, we can reasonably expect ${\bf n}_j$ to be a slowly varying
function of $j$.\\
({\em iii\/}) Associated with each ${\bf n}_j$, define a spinor
$z_{ja}$ by (\ref{nz}).\\
({\em iv\/}) With each link of the cubic lattice, we use
(\ref{Az}) to associate with it a $\mathcal{A}_{j\mu} \equiv
\mathcal{A}_{j,j+\mu}$. Here $\mu=x,y,\tau$ extends over the
3 spacetime directions.\\
With these preliminaries in hand, we can motivate the following
effective action for fluctuations under the Hamiltonian $H_s$:
\begin{eqnarray}
\mathcal{\widetilde{Z}} = \prod_{ja}\int  dz_{ja} \prod_j \delta
\left( \left| z_{ja} \right|^2 - 1 \right) \exp \left(
\frac{1}{\widetilde{g}} \sum_{\langle ij \rangle} {\bf n}_i \cdot
{\bf n}_j + i 2S \sum_j \eta_j \mathcal{A}_{j \tau} \right).
\label{zt}
\end{eqnarray}
Here the summation over $\langle ij \rangle$ extends over nearest
neighbors on the cubic lattice. The integrals are over the
$z_{ja}$, and the ${\bf n}_j$ and $\mathcal{A}_{j\tau}$ are {\em
dependent\/} variables defined via (\ref{nz}) and (\ref{Az}). Note
that both terms in the action are invariant under the gauge
transformation (\ref{gauge1}); consequently, we could equally well
have rewritten $\mathcal{\widetilde{Z}}$ as an integral over the
${\bf n}_j$, but it turns out to be more convenient to use the
$z_{ja}$ and to integrate over the redundant gauge degree of
freedom. The first term in the action contains the energy of the
Hamiltonian $H_s$, and acts to prefer nearest neighbor ${\bf n}_j$
which are parallel to each other---this ``ferromagnetic'' coupling
between the ${\bf n}_j$ in spacetime ensures, via (\ref{Sn}), that
the local quantum spin configurations are as in the N\'{e}el
state. The second term in the action is simply the Berry phase
required in the coherent state path integral, as obtained from
(\ref{single}) and (\ref{e1}): the additional factor of $\eta_j$
compensates for that in (\ref{Sn}). The dimensionless coupling
$\widetilde{g}$ controls the strength of the local
antiferromagnetic correlations; it is like a ``temperature'' for
the ferromagnet in spacetime. So for small $\widetilde{g}$ we
expect $\mathcal{\widetilde{Z}}$ to be in the N\'{e}el phase,
while for large $\widetilde{g}$ we can expect a
quantum-``disordered'' paramagnet. For a much more careful
derivation of the partition function $\mathcal{\widetilde{Z}}$
from the underlying antiferromagnet $H_s$, including a
quantitative estimate of the value of $\widetilde{g}$, see {\em
e.g.} Chapter 13 of Ref.~\cite{book}.

While it is possible to proceed with the remaining analysis of
this section using $\mathcal{\widetilde{Z}}$, we find it more
convenient to work with a very closely related alternative model.
Our proposed theory for the quantum fluctuating antiferromagnet in
its final form is \cite{SJ,sp}
\begin{eqnarray}
\mathcal{Z} &=& \prod_{j\mu} \int_0^{2 \pi}  \frac{dA_{j\mu}}{2
\pi} \prod_{ja} \int  dz_{ja} \prod_j \delta \left( \left| z_{ja}
\right|^2 - 1 \right) \nonumber \\ &~&~~~~~~~ \exp \left(
\frac{1}{g} \sum_{j\mu} \left(z^{\ast}_{ja} e^{-iA_{j\mu}}
z_{j+\mu,a} + \mbox{c.c.} \right) + i 2S \sum_j \eta_j A_{j \tau}
\right). \label{z}
\end{eqnarray}
Note that we have introduced a new field $A_{j\mu}$, on each link
of the cubic lattice, which is integrated over. Like
$\mathcal{A}_{i\mu}$, this is also a U(1) gauge field because all
terms in the action above are invariant under the analog of
(\ref{gauge2}):
\begin{equation}
A_{j\mu} \rightarrow A_{j\mu} - \phi_j + \phi_{j+\mu}.
\label{gauge3}
\end{equation}

The very close relationship between $\mathcal{Z}$ and
$\mathcal{\widetilde{Z}}$ may be seen \cite{SJ} by explicitly
integrating over the $A_{j\mu}$ in (\ref{z}): this integral can be
done exactly because the integrand factorizes into terms on each
link that depend only on a single $A_{j\mu}$. After inserting
(\ref{zzA}) into (\ref{z}), the integral over the $j\mu$ link is
\begin{eqnarray}
&& \int_0^{2 \pi} \frac{dA_{j\mu}}{2 \pi} \exp\left( \frac{(2(1+
{\bf n}_j \cdot {\bf n}_{j+\mu}))^{1/2}}{g}
\cos(\mathcal{A}_{j\mu} - A_{j\mu}) + i 2 S\eta_j
\delta_{\mu\tau} A_{j \mu} \right) \nonumber \\
&&~~~~~~~~~~~~~~~~~~=I_{2S\delta_{\mu\tau}} \left[  \frac{(2(1+
{\bf n}_j \cdot {\bf n}_{j+\mu}))^{1/2}}{g} \right] \exp\left(  i
2S \eta_j \delta_{\mu\tau} \mathcal{A}_{j \mu} \right) \label{int}
\end{eqnarray}
where the result involves either the modified Bessel function
$I_0$ (for $\mu=x,y$) or $I_{2S}$ (for $\mu = \tau$). We can use
the identity (\ref{int}) to perform the integral over $A_{j \mu}$
on each link of (\ref{z}), and so obtain a partition function,
denoted $\mathcal{Z}'$, as an integral over the $z_{ja}$ only.
This partition function $\mathcal{Z}'$ has essentially the same
structure as $\mathcal{\widetilde{Z}}$ in (\ref{zt}). The Berry
phase term in $\mathcal{Z}'$ is identical to that in
$\mathcal{\widetilde{Z}}$. The integrand of $\mathcal{Z}'$ also
contains a real action expressed solely as a sum over functions of
${\bf n}_i \cdot {\bf n}_j$ on nearest neighbor links: in
$\mathcal{\widetilde{Z}}$ this function is simply ${\bf n}_i \cdot
{\bf n}_j /\widetilde{g}$, but the corresponding function obtained
from (\ref{z}) is more complicated (it involves the logarithm of a
Bessel function), and has distinct forms on spatial and temporal
links. We do not expect this detailed form of the real action
function to be of particular importance for universal properties:
the initial simple nearest-neighbor ferromagnetic coupling between
the ${\bf n}_j$ in (\ref{zt}) was chosen arbitrarily anyway. So we
may safely work with the theory $\mathcal{Z}$ in (\ref{z})
henceforth.

One of the important advantages of (\ref{z}) is that we no longer
have to keep track of the complicated non-linear constraints
associated with (\ref{nz}) and (\ref{Az}); this was one of the
undesirable features of (\ref{zt}). In $\mathcal{Z}$, we simply
have free integration over the independent variables $z_{ja}$ and
$A_{j\mu}$. The remainder of this section will be devoted to
describing the properties of $\mathcal{Z}$ as a function of the
coupling $g$.

The theory $\mathcal{Z}$ in (\ref{z}) has some resemblance to the
so-called CP$^{N-1}$ model from the particle physics literature
\cite{dadda,witten,berg}: our indices $a,b$ take only 2 possible
values, but the general model is obtained when $a,b=1\ldots N$,
and we will also find it useful to consider $\mathcal{Z}$ for
general $N$. The case of general $N$ describes SU($N$) and Sp($N$)
antiferromagnets on the square lattice \cite{rs}. Note also that
it is essential for our purposes that the theory is invariant
under $A_{j\mu} \rightarrow A_{j\mu} + 2 \pi$, and so the U(1)
gauge theory is {\em compact}. Finally our model contains a Berry
phase term (which can be interpreted as a $J_{\mu} A_{\mu}$ term
associated with a current $J_{j\mu} = 2S \eta_j \delta_{\mu\tau} $
of static charges $\pm 2S$ on each site) which is not present in
any of the particle physics analyses. This Berry phase term will
be an essential central actor in all of our results below for the
paramagnetic phase and the quantum phase transition.

The properties of $\mathcal{Z}$ are quite evident in the limit of
small $g$. Here, the partition function is strongly dominated by
configurations in which the real part of the action is a minimum.
In a suitable gauge, these are the configurations in which $z_{ja}
= \mbox{constant}$, and by (\ref{nz}), we also have ${\bf n}_j$ a
constant. This obviously corresponds to the N\'{e}el phase with
(\ref{neel}). A Gaussian fluctuation analysis about such a
constant saddle point is easily performed, and we obtain the
expected spectrum of a doublet of gapless spin waves.

The situation is much more complicated for large $g$ where we
should naturally expect a paramagnetic phase with $\langle {\bf
S}_j \rangle = \langle {\bf n}_j \rangle = 0$. This will be
discussed in some detail in Section~\ref{sec:para}. Finally, we
will address the nature of the quantum phase transition between
the N\'{e}el and paramagnetic phases in Section~\ref{sec:qpt}.

\subsection{Paramagnetic phase}
\label{sec:para}

The discussion in this section has been adapted from another
recent review by the author \cite{th2002}.

For large $g$, we can perform the analog of a `high temperature'
expansion of $\mathcal{Z}$ in (\ref{z}). We expand the integrand
in powers of $1/g$ and perform the integral over the $z_{ja}$
term-by-term. The result is then an effective theory for the
compact U(1) gauge field $A_{j\mu}$ alone. An explicit expression
for the effective action of this theory can be obtained in powers
of $1/g$: this has the structure of a strong coupling expansion in
lattice gauge theory, and higher powers of $1/g$ yield terms
dependent upon gauge-invariant U(1) fluxes on loops of all sizes
residing on the links of the cubic lattice. For our purposes, it
is sufficient to retain only the simplest such term on elementary
square plaquettes, yielding the partition function
\begin{equation}
\mathcal{\widetilde{Z}}_A = \prod_{j\mu} \int_0^{2 \pi} \frac{d
A_{j \mu}}{2 \pi} \exp \left( \frac{1}{e^2} \sum_{\Box}
\cos\left(\epsilon_{\mu\nu\lambda}\Delta_{\nu} A_{j \lambda}
\right) - i 2S \sum_j \eta_j A_{j\tau} \right), \label{f5}
\end{equation}
where $\epsilon_{\mu\nu\lambda}$ is the totally antisymmetric
tensor in three spacetime dimensions. Here the cosine term
represents the conventional Maxwell action for a compact U(1)
gauge theory: it is the simplest local term consistent with the
gauge symmetry (\ref{gauge3}) and which is periodic under $A_{j
\mu} \rightarrow A_{j \mu} + 2 \pi$; closely related terms appear
under the $1/g$ expansion. The sum over $\Box$ in (\ref{f5})
extends over all plaquettes of the cubic lattice, $\Delta_{\mu}$
is the standard discrete lattice derivative ($\Delta_\mu f_j
\equiv f_{j+\mu} - f_j$ for any $f_j$), and $e^2$ is a coupling
constant. We expect the value of $e$ to increase monotonically
with $g$.

As is standard in duality mappings, we first rewrite the partition
function in $2+1$ spacetime dimensions by replacing the cosine
interaction in (\ref{f5}) by a Villain sum \cite{villain,jkkn}
over periodic Gaussians:
\begin{eqnarray}
\mathcal{Z}_A &=& \sum_{\{q_{\bar{\jmath}\mu}\}} \prod_{j\mu}
\int_{0}^{2\pi} \frac{d A_{j \mu}}{2 \pi} \exp \Biggl(
-\frac{1}{2e^2} \sum_{\Box} \left( \epsilon_{\mu\nu\lambda}
\Delta_{\nu} A_{j \lambda} - 2 \pi q_{\bar{\jmath}\mu} \right)^2
\nonumber \\
&~&~~~~~~~~~~~~~~~~~~~~~~~~~~~~~~~~~~~~~~~ - i 2 S \sum_j \eta_j
A_{j\tau} \Biggr), \label{f6}
\end{eqnarray}
where the $q_{\bar{\jmath}\mu}$ are integers on the links of the
{\em dual} cubic lattice, which pierce the plaquettes of the
direct lattice. Throughout this article we will use the index
$\bar{\jmath}$ to refer to sites of this dual lattice, while $j$
refers to the direct lattice on sites on which the spins are
located.

\begin{figure}[t]
\centerline{\includegraphics[width=2in]{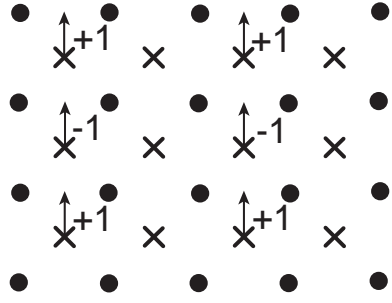}}
\caption{Specification of the non-zero values of the fixed field
$a_{\bar{\jmath}\mu}^0$. The circles are the sites of the direct
lattice, $j$, while the crosses are the sites of the dual lattice,
$\bar{\jmath}$; the latter are also offset by half a lattice
spacing in the direction out of the paper (the $\mu = \tau$
direction). The $a_{\bar{\jmath}\mu}^0$ are all zero for
$\mu=\tau,x$, while the only non-zero values of
$a_{\bar{\jmath}y}^0$ are shown above. Notice that the $a^0$ flux
obeys (\protect\ref{d2}).}\label{fig8}
\end{figure}

We will now perform a series of exact manipulations on (\ref{f6})
which will lead to a dual {\em interface} model
\cite{rs,rsb,fradkiv}. This dual model has only positive
weights---this fact, of course, makes it much more amenable to a
standard statistical analysis. This first step in the duality
transformation is to rewrite (\ref{f6}) by the Poisson summation
formula:
\begin{eqnarray}
\sum_{\{q_{\bar{\jmath}\mu}\}} \exp && \left( -\frac{1}{2e^2}
\sum_{\Box} \left(\epsilon_{\mu\nu\lambda} \Delta_{\nu} A_{j
\lambda} - 2 \pi q_{\bar{\jmath}\mu} \right)^2
\right) \nonumber \\
&&~~~~~~~~~~= \sum_{\{a_{\bar{\jmath}\mu}\}} \exp \left( -
\frac{e^2}{2} \sum_{\bar{\jmath}} a_{\bar{\jmath}\mu}^2 - i
\sum_{\Box} \epsilon_{\mu\nu\lambda} a_{\bar{\jmath}\mu}
\Delta_{\nu} A_{j \lambda}\right), \label{d1}
\end{eqnarray}
where $a_{\bar{\jmath}\mu}$ (like $q_{\bar{\jmath}\mu}$) is an
integer-valued vector field on the links of the dual lattice
(here, and below, we drop overall normalization factors in front
of the partition function). Next, we write the Berry phase in a
form more amenable to duality transformations. Choose a
`background' $a_{\bar{\jmath} \mu}=a_{\bar{\jmath} \mu}^0$ flux
which satisfies
\begin{equation}
\epsilon_{\mu\nu\lambda} \Delta_{\nu} a_{\bar{\jmath}\lambda}^0 =
\eta_j \delta_{\mu \tau}, \label{d2}
\end{equation}
where $j$ is the direct lattice site in the center of the
plaquette defined by the curl on the left-hand-side. Any
integer-valued solution of (\ref{d2}) is an acceptable choice for
$a_{\bar{\jmath}\mu}^0$, and a convenient choice is shown in
Fig~\ref{fig8}.
Using (\ref{d2}) to rewrite the Berry phase in (\ref{f6}),
applying (\ref{d1}), and shifting $a_{\bar{\jmath}\mu}$ by the
integer $2S a_{\bar{\jmath}\mu}^0$, we obtain a new exact
representation of $\mathcal{Z}_A$ in (\ref{f6}):
\begin{eqnarray}
\mathcal{Z}_A &=& \sum_{\{ a_{\bar{\jmath} \mu} \}}  \prod_{j\mu}
\int_{0}^{2\pi} \frac{d A_{j \mu}}{2 \pi} \exp \left(
-\frac{e^2}{2} \sum_{\bar{\jmath},\mu} (a_{\bar{\jmath}\mu}-2S
a_{\bar{\jmath}\mu}^0)^2 \right. \nonumber \\
&~&~~~~~~~~~~~~~~~~~~~~~~~~~~~~~~~~~~~~~\left. - i \sum_{\Box}
\epsilon_{\mu\nu\lambda} a_{\bar{\jmath}\mu} \Delta_{\nu} A_{j
\lambda} \right). \label{d4}
\end{eqnarray}
The integral over the $A_{j \mu}$ can be performed independently
on each link, and its only consequence is the imposition of the
constraint $\epsilon_{\mu\nu\lambda} \Delta_{\nu}
a_{\bar{\jmath}\lambda}=0$. We solve this constraint by writing
$a_{\bar{\jmath} \mu}$ as the gradient of a integer-valued
`height' $h_{\bar{\jmath}}$ on the sites of the dual lattice, and
so obtain
\begin{equation}
\mathcal{Z}_h = \sum_{\{ h_{\bar{\jmath}} \}} \exp \left(
-\frac{e^2}{2} \sum_{\bar{\jmath},\mu} (\Delta_{\mu}
h_{\bar{\jmath}}-2S a_{\bar{\jmath}\mu}^0)^2  \right). \label{d5}
\end{equation}
We emphasize that, apart from an overall normalization, we have
$\mathcal{Z}_h = \mathcal{Z}_A$ exactly. This is the promised 2+1
dimensional interface, or height, model in almost its final form.

\begin{figure}[t]
\centerline{\includegraphics[width=4in]{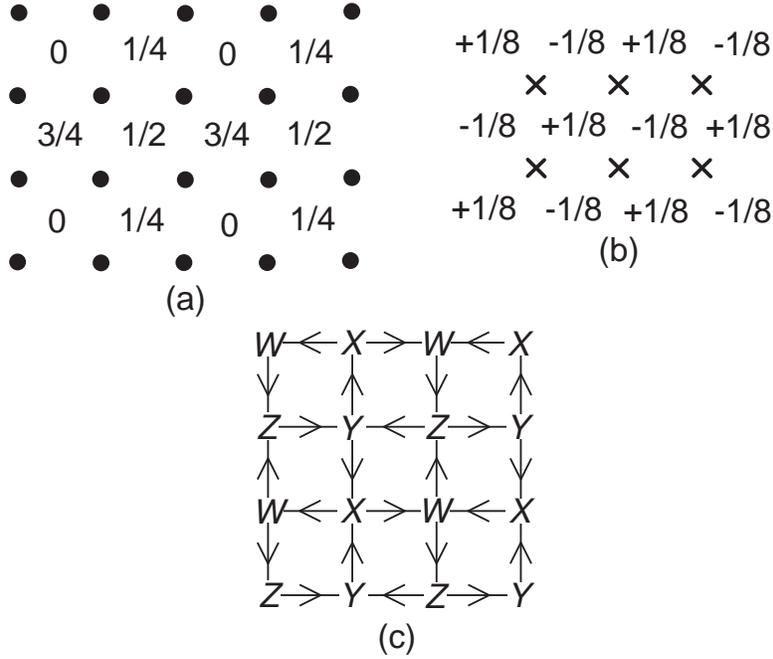}}
\caption{Specification of the non-zero values of the fixed fields
(a) $\mathcal{X}_{\bar{\jmath}}$, (b) $\mathcal{Y}_{j \mu}$, (c)
$\epsilon_{\mu\nu\lambda} \Delta_{\nu} \mathcal{Y}_{j \lambda}$
introduced in (\protect\ref{XY}). The notational conventions are
as in Fig~\protect\ref{fig8}. Only the $\mu=\tau$ components of
$\mathcal{Y}_{j \mu}$ are non-zero, and these are shown in (b).
Only the spatial components of $\epsilon_{\mu\nu\lambda}
\Delta_{\nu} \mathcal{Y}_{j \lambda}$ are non-zero, and these are
oriented as in (c) with magnitude 1/4. The four dual sublattices,
$W$, $X$, $Y$, $Z$, are also indicated in (c). Note that
$\mathcal{X}_W = 0$, $\mathcal{X}_X = 1/4$, $\mathcal{X}_Y = 1/2$,
and $\mathcal{X}_Z = 3/4$.}\label{fig9}
\end{figure}

The physical properties of (\ref{d5}) become clearer by converting
the ``frustration'' $a_{\bar{\jmath}\mu}^0$ in (\ref{d5}) into
offsets for the allowed height values. This is done by decomposing
$a_{\bar{\jmath}\mu}^0$ into curl and divergence free parts and
writing it in terms of new fixed fields,
$\mathcal{X}_{\bar{\jmath}}$ and ${\mathcal Y}_{j \mu}$ as
follows:
\begin{equation}
a_{\bar{\jmath}\mu}^{0} = \Delta_{\mu} \mathcal{X}_{\bar{\jmath}}
+ \epsilon_{\mu\nu\lambda} \Delta_{\nu} \mathcal{Y}_{j  \lambda}.
\label{XY}
\end{equation}
The values of these new fields are shown in Fig~\ref{fig9}.
Inserting (\ref{XY}) into (\ref{d5}), we can now write the height
model in its simplest form \cite{rsb}
\begin{equation}
\mathcal{Z}_h = \sum_{\{H_{\bar{\jmath}}\}} \exp \left ( -
\frac{e^2}{2} \sum_{\bar{\jmath}} \left( \Delta_{\mu}
H_{\bar{\jmath}} \right)^2 \right), \label{he1}
\end{equation}
where
\begin{equation}
H_{\bar{\jmath}} \equiv h_{\bar{\jmath}} - 2 S
\mathcal{X}_{\bar{\jmath}} \label{he2}
\end{equation}
is the new height variable we shall work with. Notice that the
$\mathcal{Y}_{j \mu}$ have dropped out, while the
$\mathcal{X}_{\bar{\jmath}}$ act only as fractional offsets (for
$S$ not an even integer) to the integer heights. From (\ref{he2})
we see that for half-odd-integer $S$ the height is restricted to
be an integer on one of the four sublattices, an integer plus 1/4
on the second, an integer plus 1/2 on the third, and an integer
plus 3/4 on the fourth; the fractional parts of these heights are
as shown in Fig~\ref{fig9}a; the steps between neighboring heights
are always an integer plus 1/4, or an integer plus 3/4. For $S$ an
odd integer, the heights are integers on one square sublattice,
and half-odd-integers on the second sublattice. Finally for even
integer $S$ the offset has no effect and the height is an integer
on all sites. We discuss these classes of $S$ values in turn in
the following subsections.

\subsubsection{4.1.1~$S$ even integer} \label{S2}

In this case the offsets $2S \mathcal{X}_{\bar{\jmath}}$ are all
integers, and (\ref{he1}) is just an ordinary three dimensional
height model which has been much studied in the literature
\cite{jkkn,rough}. Unlike the two-dimensional case,
three-dimensional height models generically have no roughening
transition, and the interface is always smooth \cite{rough}. With
all heights integers, the smooth phase breaks no lattice
symmetries. So square lattice antiferromagnets with $S$ even
integer can have a paramagnetic ground state with a spin gap and
no broken symmetries. The smooth interface corresponds to
confinement in the dual compact U(1) gauge theory \cite{polyakov}:
consequently the $z_a$ of $\mathcal{Z}$ are confined, and the
elementary excitations are $S=1$ quasiparticles, similar to the
$\varphi_\alpha$ of $\mathcal{S}_\varphi$. This is in accord with
the exact ground state for a $S=2$ antiferromagnet on the square
lattice found by Affleck {\em et al.}, the AKLT state \cite{aklt}.

\subsubsection{4.1.2~$S$ half-odd-integer} \label{Shint}

Now the heights of the interface model can take four possible
values, which are integers plus the offsets on the four square
sublattices shown in Fig~\ref{fig9}a. As in Section~\ref{S2}.1,
the interface is always smooth {\em i.e.} any state of (\ref{he1})
has a fixed average interface height
\begin{equation}
\overline{H} \equiv \frac{1}{N_d} \sum_{\bar{\jmath}=1}^{N_d}
\langle H_{\bar{\jmath}} \rangle,
\end{equation}
where the sum is over a large set of $N_d$ dual lattice points
which respect the square lattice symmetry. {\em Any} well-defined
value for $\overline{H}$ breaks the uniform shift symmetry of the
height model under which $H_{\bar{\jmath}} \rightarrow
H_{\bar{\jmath}} \pm 1$. In the present context, only the value of
$\overline{H}$ modulo integers is physically significant, and so
the breaking of the shift symmetry is not important by itself.
However, after accounting for the height offsets, we now prove
that {\em any smooth interface must also break a lattice symmetry
with the development of bond order}: this means that
$\mathcal{Z}_A$ in (\ref{f6}) describes spin gap ground states of
the lattice antiferromagnet which necessarily have spontaneous
bond order.

The proof of this central result becomes clear upon a careful
study of the manner in which the height model in (\ref{he1}) and
(\ref{he2}) implements the 90$^\circ$ rotation symmetry about a
direct square lattice point. Consider such a rotation under which
the dual sublattice points in Fig~\ref{fig9}c interchange as
\begin{equation}
W \rightarrow X,~~ X \rightarrow Y,~~ Y \rightarrow Z,~~ Z
\rightarrow W. \label{hf1}
\end{equation}
The terms in the action in (\ref{he2}) will undergo a 90$^\circ$
rotation under this transformation provided the integer heights
$h_{\bar{\jmath}}$ transform as
\begin{equation}
h_W \rightarrow h_X,~~ h_X \rightarrow h_Y,~~ h_Y \rightarrow
h_Z,~~ h_Z \rightarrow h_W - 1. \label{hf2}
\end{equation}
Notice the all important $-1$ in the last term---this compensates
for the `branch cut' in the values of the offsets
$\mathcal{X}_{\bar{\jmath}}$ as one goes around a plaquette in
Fig~\ref{fig9}c. From (\ref{hf2}), it is evident that the average
height $\overline{H} \rightarrow \overline{H} - 1/4$ under the
90$^\circ$ rotation symmetry under consideration here. Hence, a
smooth interface with a well-defined value of $\overline{H}$ {\em
always} breaks this symmetry.

\begin{figure}[t]
\centerline{\includegraphics[width=2in]{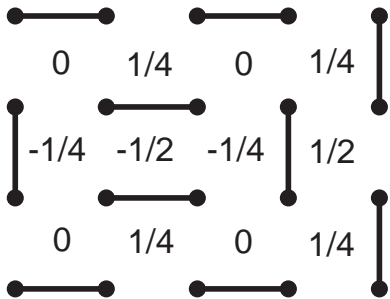}}
\caption{Mapping between the quantum dimer model and the interface
model $\mathcal{Z}_h$ in (\protect\ref{he1}). Each dimer on the
direct lattice is associated with a step in height of $\pm 3/4$ on
the link of the dual lattice that crosses it. All other height
steps are $\pm 1/4$. Each dimer represents a singlet valence bond
between the sites, as in Fig~\protect\ref{fig2}.}\label{fig10}
\end{figure}

We now make this somewhat abstract discussion more physical by
presenting a simple interpretation of the interface model in the
language of the $S=1/2$ antiferromagnet \cite{zheng}. From
Fig~\ref{fig9}a it is clear that nearest neighbor heights can
differ either by 1/4 or 3/4 (modulo integers). To minimize the
action in (\ref{he1}), we should choose the interface with the
largest possible number of steps of $\pm 1/4$. However, the
interface is frustrated, and it is not possible to make all steps
$\pm 1/4$ and at least a quarter of the steps must be $\pm 3/4$.
Indeed, there is a precise one-to-one mapping between interfaces
with the minimal number of $\pm 3/4$ steps (we regard interfaces
differing by a uniform integer shift in all heights as equivalent)
and the dimer coverings of the square lattice: the proof of this
claim is illustrated in Fig~\ref{fig10}.
We identify each dimer with a singlet valence bond between the
spins (the ellipses in Fig~\ref{fig2}), and so each interface
corresponds to a quantum state with each spin locked in a singlet
valence bond with a particular nearest neighbor. Fluctuations of
the interface in imaginary time between such configurations
correspond to quantum tunneling events between such dimer states,
and an effective Hamiltonian for this is provided by the quantum
dimer model \cite{qd}. While such an interpretation in terms of
the dimer model is appealing, we should also note that it is not
as general as the dual interface model: on certain lattices, while
the collinear paramagnetic state continues to have a
representation as a dual interface model, there is no
corresponding dimer interpretation \cite{kimchung}.

\begin{figure}[t]
\centerline{\includegraphics[width=2.7in]{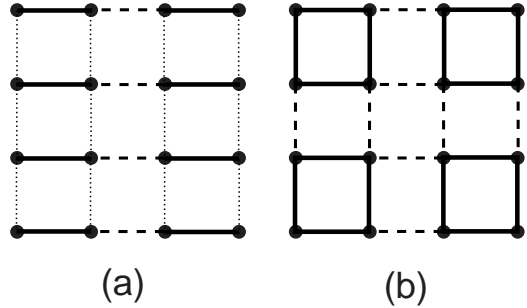}}
\caption{Sketch of the two simplest possible states with bond
order for $S=1/2$ on the square lattice: (a) the columnar
spin-Peierls states, and (b) plaquette state. Here the distinct
line styles encode the different values of the bond order
parameter $Q_{ij}$ in (\protect\ref{bondorder}) on the links. This
should be contrasted from
Figs.~\protect\ref{fig1}-\protect\ref{fig4} where the line styles
represented distinct values of the exchange constants in the
Hamiltonian. In the present section, the Hamiltonian has the full
symmetry of the square lattice, and the orderings represented
above amount to a spontaneous breaking of the lattice symmetry.
Both states above are 4-fold degenerate; an 8-fold degenerate
state, with superposition of the above orders, also appears as a
possible ground state of the generalized interface model.
Numerical studies of a number of two-dimensional quantum
antiferromagnets
\protect\cite{fouet1,brenig,fouet2,sandvik,harada,kotov,van} have
found ground states with spontaneous bond order, similar to the
states shown above.} \label{fig11}
\end{figure}

The nature of the possible smooth phases of the interface model
are easy to determine from the above picture and by standard
techniques from statistical theory \cite{rsb,zheng}. As a simple
example, the above mapping between interface heights and dimer
coverings allows one to deduce that interfaces with average height
$\overline{H} = 1/8,3/8,5/8,7/8$ (modulo integers) correspond to
the four-fold degenerate bond-ordered states in Fig~\ref{fig11}a.
To see this, select the interface with $h_{\bar{\jmath}} = 0$ for
all $\bar{\jmath}$: this interface has the same symmetry as
Fig~\ref{fig11}a, and a simple computation summing over sites from
(\ref{he2}) shows that this state has average height $\overline{H}
= -(0+1/4+1/2+3/4)/4=-3/8$ for $S=1/2$. The remaining three values
of $\overline{H}$ correspond to the three other states obtained by
successive 90$^\circ$ rotations of Fig~\ref{fig11}a. In a similar
manner, interfaces with $\overline{H} = 0,1/4,1/2,3/4$ (modulo
integers) correspond to the four-fold degenerate plaquette
bond-ordered states in Fig~\ref{fig11}b. A simple example of such
an interface is the ``disordered-flat'' state \cite{disflat} in
which $h_{\bar{\jmath}}=0$ on all sites $\bar{\jmath}$, except for
the $W$ sublattice which have $\mathcal{X}_{\bar{\jmath}}=0$; for
these sites we have $h_{\bar{\jmath}}$ fluctuate randomly between
$h_{\bar{\jmath}}=0$ and $h_{\bar{\jmath}}=1$, and independently
for different $\bar{\jmath}$. The average height of such an
interface is $\overline{H} = -((0+1)/2+1/4+1/2+3/4)/4= -1/2$ for
$S=1/2$, and the mapping to dimer coverings in Fig~\ref{fig10}
shows easily that such an interface corresponds to the state in
Fig~\ref{fig11}b. All values of $\overline{H}$ other than those
quoted above are associated with eight-fold degenerate
bond-ordered states with a superposition of the orders in
Fig~\ref{fig11}a and b.

All these phases are expected to support non-zero spin
quasiparticle excitations which carry spin $S=1$, but not $S=1/2$.
Despite the local corrugation in the interface configuration
introduced by the offsets, the interface remains smooth on the
average, and this continues to correspond to confinement in the
dual compact U(1) gauge theory \cite{polyakov}. Consequently the
spinons of Fig~\ref{fig3}b are confined in pairs. The structure of
the resulting $S=1$ triplon quasiparticles is very similar to the
excitations of the paramagnetic phase of the coupled dimer
antiferromagnet of Section~\ref{sec:dimer}, as we already noted in
Section~\ref{sec:intro}.

Support for the class of bond-ordered states described above has
appeared in a number of numerical studies of $S=1/2$
antiferromagnets in $d=2$ which have succeeded in moving from the
small $g$ N\'{e}el phase to the large $g$ paramagnet. These
include studies on the honeycomb lattice \cite{fouet1} (duality
mapping on the honyecomb lattice appears in Ref.~\cite{rs}), on
the planar pyrochlore lattice \cite{brenig,fouet2} (duality
mapping for a lattice with the symmetry of the planar pyrochore is
in Refs.~\cite{chung,kimchung}, with a prediction for the bond
order observed), on square lattice models with ring-exchange and
easy-plane spin symmetry \cite{sandvik} (duality mapping on spin
models with easy plane symmetry is in Refs.~\cite{lfs,sp,kwon}),
and square lattice models with SU($N$) symmetry \cite{harada} (the
theories (\ref{z}), with $a=1\ldots N$, and (\ref{he1}) apply
unchanged to SU($N$) antiferromagnets). The case of the square
lattice antiferromagnet with first and second neighbor exchange is
not conclusively settled: while two recent studies
\cite{kotov,van} (and earlier work \cite{gsh,ziman}) do observe
bond order in a paramagnetic spin-gap state, a third
\cite{sorella} has so far not found such order. It is possible
that this last study is observing signatures of the critical point
between the N\'{e}el and bond-ordered states (to be described in
Section~\ref{sec:qpt}) which is expressed in a theory for
deconfined spinons in $\mathcal{Z}_c$ in (\ref{zc}).

Finally, we also mention that evidence for the spontaneous bond
order of Fig~\ref{fig11} appears in recent numerical studies of
{\em doped\/} antiferromagnets \cite{sushkovd,eder}.

\subsubsection{4.1.3~$S$ odd integer}

This case is similar to that $S$ half-odd-integer, and we will not
consider it in detail. The Berry phases again induce bond order in
the spin gap state, but this order need only lead to a two-fold
degeneracy.

\subsection{Critical theory}
\label{sec:qpt}

We turn finally to the very difficult issue of the nature of the
quantum phase transition from the N\'{e}el state to one of the
bond-ordered paramagnetic states in Fig~\ref{fig10} as a function
of increasing $g$. This has been a long-standing open problem, and
many different proposals have been made. The two phases break
different symmetries of the Hamiltonian, and so are characterized
by very different order parameters (one lives in spin space, and
the other in real space). Landau-Ginzburg-Wilson (LGW) theory
would imply that a generic second-order transition is not possible
between such phases, and one obtains either a first-order
transition  or a region of co-existence of the two orders.
However, the bond-order in the paramagnet was obtained entirely
from quantum Berry phases attached to the fluctuating N\'{e}el
order, and it is not clear that LGW theory applies in such a
situation.

Recent work by Senthil {\em et al.} \cite{senthil,glw} has
proposed an elegant resolution to many of these problems, and we
will describe their results in the remainder of this subsection.
The results are based upon solutions of a series of simpler models
which strongly suggest that related results also apply to the
SU(2) invariant, $S=1/2$ models of interest. The computations are
intricate, but the final results are quite easy to state, and are
presented below. We will mainly limit our discussion here to the
case of antiferromagnets of spin $S=1/2$.

First, contrary to the predictions of LGW theory, a generic
second-order transition between the N\'{e}el state and the
bond-ordered paramagnet is indeed possible (let us assume it
occurs at $g=g_c$ for $\mathcal{Z}$ in (\ref{z})). The theory for
such a quantum critical point is obtained simply by taking a naive
continuum limit of $\mathcal{Z}$ while ignoring {\em both} the
compactness of the gauge field and the Berry phases. Remarkably,
these complications of the lattice model $\mathcal{Z}$, which we
have so far stated were essential for the complete theory, have
effects which cancel each other out, but {\em only} at the
critical point. Note compactness on its own is a relevant
perturbation which cannot be ignored {\em i.e.} without Berry
phases, the compact and non-compact lattice CP$^1$ model have
distinct critical theories \cite{mv}. However, the surprising new
point noted by Senthil {\em et al.} \cite{senthil,glw} is that the
{\em non-compact CP$^1$ model has the same critical theory as the
compact CP$^1$ model with $S=1/2$ Berry phases}. Taking the naive
continuum limit of $\mathcal{Z}$ in (\ref{z}), and softening the
hard-constraint on the $z_{ja}$, we obtain the proposed theory for
the quantum critical point between the N\'{e}el state and the
bond-ordered paramagnet for spin $S=1/2$\cite{senthil,glw}:
\begin{eqnarray}
\mathcal{Z}_c &=& \int \mathcal{D} z_a (r, \tau) \mathcal{D}
A_{\mu} (r, \tau) \exp \Biggl( - \int d^2 r d \tau \biggl[
|(\partial_\mu -
i A_{\mu}) z_a |^2 + s |z_a |^2  \nonumber \\
&~&~~~~~~~~~~~~~~~~~~~~~ ~~~~ + \frac{u}{2} (|z_a|^2)^2 +
\frac{1}{4e^2} (\epsilon_{\mu\nu\lambda}
\partial_\nu A_\lambda )^2 \biggl]\Biggl) \label{zc}
\end{eqnarray}
We have also included here a kinetic term for the $A_{\mu}$, and
one can imagine that this is generated by integrating out large
momentum $z_{ja}$. On its own, $\mathcal{Z}_c$ describes the
transition from a magnetically ordered phase with $z_a$ condensed
at $s<s_c$, to a disordered state with a gapless U(1) photon at $s
>s_c$ (here $s_c$ is the critical point of $\mathcal{Z}_c$).
Clearly the $s<s_c$ phase corresponds to the N\'{e}el phase of
$\mathcal{Z}$ in (\ref{z}) for $g < g_c$. However, the $s>s_c$
phase does not obviously correspond to the $g>g_c$ bond-ordered,
fully gapped, paramagnet of $\mathcal{Z}$. This is repaired by
accounting for the compactness of the gauge field and the Berry
phases: it is no longer possible to neglect them, while it was
safe to do so precisely at $g=g_c$. The {\em combined} effects of
compactness and Berry phases are therefore {\em dangerously
irrelevant} at $g=g_c$.

It is important to note that the critical theory of (\ref{zc}) is
distinct from the critical theory $\mathcal{S}_{\varphi}$ in
(\ref{sp}), although both theories have a global O(3) symmetry
\cite{mv}. In particular the values of the exponents $\nu$ are
different in the two theories, and the scaling dimension of the
N\'eel order parameter $\varphi_\alpha$ under
$\mathcal{S}_\varphi$ is distinct from the scaling dimension of
the N\'eel order parameter $z_a^\ast \sigma^{\alpha}_{ab} z_b$ at
the critical point of $\mathcal{Z}_c$.

It is interesting that $\mathcal{Z}_c$ in (\ref{zc}) is a theory
for the $S=1/2$ spinors $z_a$. These can be understood to be the
continuum realization of the spinons shown earlier in
Fig~\ref{fig3}b. Thus the spinons become the proper elementary
degrees of freedom, but {\em only} at the quantum critical point.
Hence it is appropriate to label this as a `deconfined quantum
critical point' \cite{senthil}. These spinons are confined into a
$S=1$ quasiparticle once bond order appears for $g>g_c$, for
reasons similar to those illustrated in Fig~\ref{fig3}b.

A key characteristic of this `deconfined' critical point is the
irrelevance of the compactness of the gauge field, and hence of
monopole tunnelling events. A consequence of this is that the flux
of the $A_{\mu}$ gauge field in $\mathcal{Z}_c$ is conserved. This
emergent conservation law, and the associated long-range gauge
forces are key characteristics of such critical points.

\begin{figure}[t]
\centerline{\includegraphics[width=4.5in]{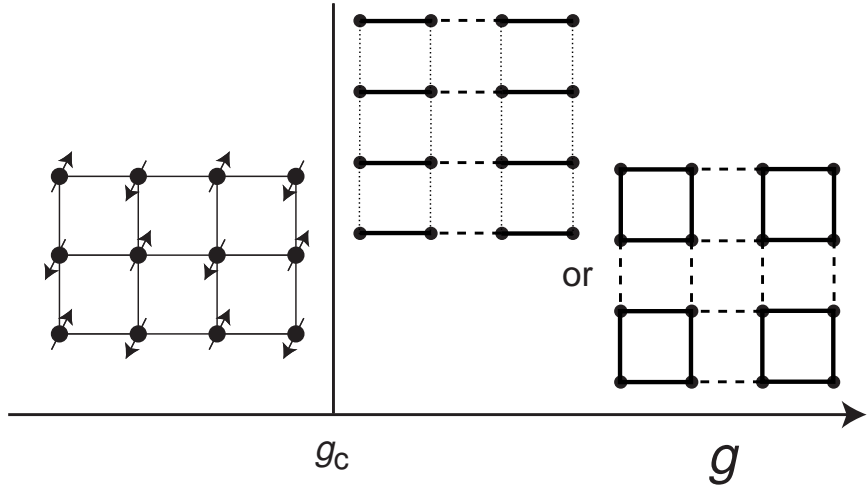}}
\caption{Phase diagram of the model $\mathcal{Z}$ in
(\protect\ref{z}) of $S=1/2$ antiferromagnets with full square
lattice symmetry. There is a N\'{e}el phase for $g<g_c$ which
breaks spin rotation invariance; it has a doublet of gapless spin
wave excitations. The bond-ordered paramagnet for $g>g_c$
preserves spin rotation invariance but breaks square lattice
symmetry; it has a gap to all excitations, and the non-zero spin
excitations are described by $S=1$ triplet quasiparticles which
are very similar to the `triplons' discussed in
Section~\ref{sec:lad1}. The critical point at $g=g_c$ is described
by the theory of $S=1/2$ `spinons', $\mathcal{Z}_c$ in
(\protect\ref{zc}) at its critical point $s=s_c$; note that this
mapping to the spinon theory $\mathcal{Z}_c$ does not work away
from $g=g_c$, and spinons are confined for all $g>g_c$. A phase
diagram like the one above has been used as a point of departure
to obtain a phase diagram for doped Mott insulators
\protect\cite{ijmp,vojta}, as a description of the cuprate
superconductors; evidence for spontaneous bond order in doped
antiferromagnets appears in Refs.~\protect\cite{sushkovd,eder}.}
\label{fig12}
\end{figure}

We summarize in Fig~\ref{fig12} our results for $S=1/2$ square
lattice antiferromagnets, as described by $\mathcal{Z}$ in
(\ref{z}).

The claims above for the conspiracy between the compactness and
Berry phases at the critical point are surprising and new. They
are central to a complete understanding of square lattice
antiferromagnets, and a full justification of the claims appears
in the work of Senthil {\em et al.}. The following subsections
illustrate their origin by considering a series of models, of
increasing complexity, where similar phenomena can be shown to
occur. The reader may also find it useful to look ahead to Tables
1 and 2, which summarize the intricate relationships between the
models considered.

\subsubsection{4.2.1~Lattice model at $N=1$}
\label{n1}

This subsection describes a simplified lattice gauge theory model
introduced by Sachdev and Jalabert \cite{SJ}. While the duality
analysis presented below was initiated in Ref.~\cite{SJ}, its
correct physical interpretation, and the implications for more
general models are due to Senthil {\em et al.} \cite{senthil,glw}.

The model of interest in this subsection is the $N=1$ case of
$\mathcal{Z}$. Physically, such a model will be appropriate for an
antiferromagnet in the presence of a staggered magnetic field:
such a field will prefer $z_{\uparrow}$ over $z_{\downarrow}$
(say). So we write the preferred single component complex scalar
simply as $z_j = e^{i \theta_j}$, and obtain from (\ref{z})
\begin{eqnarray}
\mathcal{Z}_1 &=& \prod_j \int_0^{2 \pi} \frac{d\theta_{j}}{2 \pi}
\int_0^{2 \pi} \frac{dA_{j \mu}}{2 \pi} \exp \left( \frac{1}{e^2}
\sum_{\Box} \cos \left( \epsilon_{\mu\nu\lambda}\Delta_{\nu} A_{j
\lambda}
\right) \right. \nonumber \\
&~&~~~~~~~~~~~~~~ \left. + \frac{1}{g} \sum_{j, \mu} \cos \left(
\Delta_{\mu} \theta_j - A_{j \mu} \right) + i 2S \sum_j \eta_j
A_{j \tau} \right). \label{z1}
\end{eqnarray}
We have chosen here to explicitly include a compact Maxwell term
for the gauge field, as that proves convenient in the description
of the duality mappings. Note that if we integrate out the
$\theta_j$ for large $g$, then we again obtain the model
$\mathcal{Z}_A$ in (\ref{f5}) which was used to describe the
paramagnetic phase in Section~\ref{sec:para}. So bond order
appears also in the model $\mathcal{Z}_1$ at large $g$. This bond
order disappears as $g$ is reduced, at a transition we will
describe below.

Rather than attack $\mathcal{Z}_1$ directly, it is useful as a
warm-up, and to make contact with previous work, to consider a
sequence of simpler models that have been considered in the
literature. As we have emphasized, $\mathcal{Z}_1$ features the
combined complications of compactness and Berry phases, essential
for a proper description of quantum antiferromagnets. It is the
simplest model in which it can be shown that these complications
effectively neutralize one another at the critical point.

In the following subsection, we make things simpler for ourselves
momentarily by dropping {\em both} the compactness and the Berry
phases. We will then, in the subsequent subsections, add these
complications back in.

\paragraph{A. \underline{XY model with a non-compact U(1) gauge
field}}

Dropping both compactness and Berry phases, $\mathcal{Z}_1$
reduces to
\begin{eqnarray}
\mathcal{Z}_{\rm SC} &=& \prod_j \int_0^{2 \pi}
\frac{d\theta_{j}}{2 \pi} \int_{-\infty}^{\infty} dA_{j \mu} \exp
\left( -\frac{1}{2e^2} \sum_{\Box} \left(
\epsilon_{\mu\nu\lambda}\Delta_{\nu} A_{j \lambda}
\right)^2 \right. \nonumber \\
&~&~~~~~~~~~~~~~~~~~~~~~~~~~~~~~~~~~~~~ \left.  +\frac{1}{g}
\sum_{j, \mu} \cos \left( \Delta_{\mu} \theta_j - A_{j \mu}
\right) \right). \label{zsc}
\end{eqnarray}
Notice that the Maxwell term for the gauge field now has a simple
Gaussian form. This is simply the lattice, classical,
Ginzburg-Landau model (or an XY model) of a superconductor at
finite temperatures coupled to electromagnetism. This model has
been studied extensively in the past, and the key result was
provided by Dasgupta and Halperin \cite{dh}. As we review below,
they showed that $\mathcal{Z}_{\rm SC}$ exhibited an {\em inverted
XY transition \/} {\em i.e.\/} it was dual to the theory of a
complex scalar $\psi$ in the absence of a gauge field:
\begin{equation}
\mathcal{Z}_{\rm SC,dual} = \int \mathcal{D}\psi (r, \tau) \exp
\left( - \int d^2 r d \tau \left( |\partial_{\mu} \psi|^2 +
\overline{s} |\psi|^2 + \frac{\overline{u}}{2} |\psi|^4 \right)
\right) \label{zscd}
\end{equation}
The field $\psi$ is a creation operator for {\em vortices} in the
original theory of the Ginzburg-Landau superconductor. These have
a short-range interaction ($\overline{u}$ above) because of the
screening provided by the electromagnetic flux quantum attached to
every vortex in (\ref{zsc}). So the vortex loops of (\ref{zsc})
behave like the world lines of the dual boson field of
(\ref{zscd}). The tuning parameter $\overline{s}$ in (\ref{zscd})
is `inverted' from the perspective of the direct theory: the
$\overline{s} < \overline{s}_c$ phase with  $\langle \psi \rangle
\neq 0$ has a vortex condensate and so is the normal state of a
Ginzburg-Landau superconductor, while the $\overline{s} >
\overline{s}_c$ phase with  $\langle \psi \rangle = 0$ has the
vortices gapped as in the superconducting phase.

We now provide a few steps in the analysis which links (\ref{zsc})
to (\ref{zscd}). The steps are very similar to those described in
Section~\ref{sec:para} below (\ref{f5}) and (\ref{f6}). We write
the cosine in (\ref{zsc}) in its Villain form, decouple it by the
Poisson summation formula using integer currents $J_{j\mu}$, and
also decouple the Maxwell term by a Hubbard-Stratonovich field
$P_{\bar{\jmath}\mu}$; this yields the analog of (\ref{d1}) for
$\mathcal{Z}_{\rm SC}$:
\begin{eqnarray}
&& \mathcal{Z}_{\rm SC,1} = \prod_j \int_0^{2 \pi}
\frac{d\theta_{j}}{2 \pi} \int_{-\infty}^{\infty} dA_{j \mu}
\sum_{\{J_{j\mu}\}} \int_{-\infty}^{\infty} dP_{\bar{\jmath}\mu}
\exp \left( -\frac{e^2}{2} \sum_{\bar{\jmath},\mu}
P_{\bar{\jmath}\mu}^2 \right. \nonumber \\
&&~~~~~~\left. - \frac{g}{2} \sum_{j\mu} J_{j\mu}^2 + i \sum_{j}
J_{j\mu} \left(\Delta_\mu \theta_j - A_{j\mu}\right) + i
\sum_{\Box} \epsilon_{\mu\nu\lambda} P_{\bar{\jmath} \mu}
\Delta_{\nu} A_{j \lambda} \right). \label{zsc1}
\end{eqnarray}
The advantage of this form is that the integrals over $\theta_j$
and $A_{j \mu}$ can be performed exactly, and they lead to the
constraints
\begin{equation}
\Delta_{\mu} J_{j \mu} = 0~~~~;~~~~~J_{j\mu} =
\epsilon_{\mu\nu\lambda} \Delta_{\nu} P_{\bar{\jmath} \lambda}.
\label{jp}
\end{equation}
We solve these constraints by writing
\begin{equation}
J_{j \mu} = \epsilon_{\mu\nu\lambda} \Delta_{\nu} b_{\bar{\jmath}
\lambda}~~~;~~~P_{\bar{\jmath}\mu} = b_{\bar{\jmath}\mu} -
\Delta_{\mu} \varphi_{\bar{\jmath}},
\end{equation}
where $b_{\bar{\jmath} \mu}$ is an integer valued field on the
links of the dual lattice, and $\varphi_{\bar{\jmath}}$ is a real
valued field on the sites of the dual lattice. This transforms
(\ref{zsc1}) to
\begin{eqnarray}
&& \mathcal{Z}_{\rm SC,2} = \prod_{\bar{\jmath}}
\int_{-\infty}^{\infty} d \varphi_{\bar{\jmath}}
\sum_{\{b_{\bar{\jmath}\mu}\}} \exp \left( -\frac{e^2}{2}
\sum_{\bar{\jmath},\mu} \left(b_{\bar{\jmath}\mu} -
\Delta_{\mu} \varphi_{\bar{\jmath}}\right)^2 \right. \nonumber \\
&&~~~~~~~~~~~~~~~~~~~~~~~~~~~~~~~~~~~~~~~~~~\left. - \frac{g}{2}
\sum_{\Box} \left( \epsilon_{\mu\nu\lambda} \Delta_{\nu}
b_{\bar{\jmath} \lambda} \right)^2 \right); \label{zsc2}
\end{eqnarray}
precisely this dual form was obtained by Dasgupta and Halperin
\cite{dh}, and used by them for numerical simulations. We proceed
further analytically, using methods familiar in the theory of
duality mappings \cite{jkkn}: we promote the integer valued
$b_{\bar{\jmath}\mu}$ to a real field by the Poisson summation
method, and introduce, by hand, a vortex fugacity $y_v$. This
transforms $\mathcal{Z}_{\rm SC,2}$ to
\begin{eqnarray}
&& \mathcal{Z}_{\rm SC,3} = \prod_{\bar{\jmath}}
\int_{-\infty}^{\infty} d b_{\bar{\jmath}\mu}
\int_{-\infty}^{\infty} d \varphi_{\bar{\jmath}}
\int_{-\infty}^{\infty} d \vartheta_{\bar{\jmath}} \exp \left(
-\frac{e^2}{2} \sum_{\bar{\jmath},\mu} \left(b_{\bar{\jmath}\mu} -
\Delta_{\mu} \varphi_{\bar{\jmath}}\right)^2 \right. \nonumber \\
&&~~~~~\left. - \frac{g}{2} \sum_{\Box} \left(
\epsilon_{\mu\nu\lambda} \Delta_{\nu} b_{\bar{\jmath} \lambda}
\right)^2 + y_v \sum_{\bar{\jmath},\mu} \cos \left(2 \pi
b_{\bar{\jmath}\mu} - \Delta_{\mu} \vartheta_{\bar{\jmath}}
\right) \right). \label{zsc3}
\end{eqnarray}
Notice that the effect of the vortex fugacity is to yield the
least action when $b_{\bar{\jmath}\mu}$ is an integer (ignore
$\vartheta_{\bar{\jmath}}$ momentarily): so we have effectively
`softened' the integer constraint on $b_{\bar{\jmath}\mu}$. We
have also introduced here a new real valued field
$\vartheta_{\bar{\jmath}}$ on the sites of the dual lattice simply
to make the $\mathcal{Z}_{\rm SC,3}$ invariant under U(1) gauge
transformations of $b_{\bar{\jmath}\mu}$. This is mainly because
the physics is clearer in this explicitly gauge-invariant form. We
could, if we had wished, also chosen a gauge in which
$\vartheta_{\bar{\jmath}} = 0$, and then the field
$\vartheta_{\bar{\jmath}}$ would not be present in
$\mathcal{Z}_{\rm SC,3}$ (this justifies neglect of
$\vartheta_{\bar{\jmath}}$ above). In the complete form in
(\ref{zsc3}), it is clear from the first two Gaussian terms that
fluctuations of the $b_{\bar{\jmath}\mu}$ gauge field have been
`Higgsed' by the real field $\varphi_{\bar{\jmath}}$. Indeed, it
is more convenient to choose a gauge in which
$\varphi_{\bar{\jmath}} = 0$, and we do so. Now the fluctuations
of $b_{\bar{\jmath}\mu}$ are `massive' and so can be safely
integrated out. To leading order in $y_v$, this involves simply
replacing $b_{\bar{\jmath}\mu}$ with the saddle point value
obtained from the first two Gaussian terms, which is
$\overline{b}_{\bar{\jmath}\mu}=0$. So we have the very simple
final theory
\begin{equation}
\mathcal{Z}_{\rm SC,4} = \prod_{\bar{\jmath}}
\int_{-\infty}^{\infty} d \vartheta_{\bar{\jmath}} \exp \left( y_v
\sum_{\bar{\jmath},\mu} \cos \left( \Delta_{\mu}
\vartheta_{\bar{\jmath}} \right) \right), \label{zsc4}
\end{equation}
which has the form of the dual XY model. We now take the continuum
limit of (\ref{zsc4}) by a standard procedure \cite{hbst} of
introducing a complex field $\psi$ conjugate to  $e^{i
\vartheta_{\bar{\jmath}}}$, and obtain the theory
$\mathcal{Z}_{\rm SC,dual}$ as promised. This establishes the
duality mapping of Dasgupta and Halperin \cite{dh}.

\paragraph{B. \underline{XY model with a compact U(1) gauge field}}

Now we ease towards our aim of a duality analysis of
$\mathcal{Z}_1$, by adding one layer of complexity to
$\mathcal{Z}_{\rm SC}$. We make the gauge field in (\ref{zsc})
compact by including a cosine Maxwell term \cite{sudbo}:
\begin{eqnarray}
\mathcal{Z}_M &=& \prod_j \int_0^{2 \pi} \frac{d\theta_{j}}{2 \pi}
\int_0^{2 \pi} \frac{dA_{j \mu}}{2 \pi}  \exp \left( \frac{1}{e^2}
\sum_{\Box} \cos \left( \epsilon_{\mu\nu\lambda}\Delta_{\nu} A_{j
\lambda}
\right) \right. \nonumber \\
&~&~~~~~~~~~~~~~~~~~~~~~~~~~~~~~~~~~~~~ \left.  + \frac{1}{g}
\sum_{j, \mu} \cos \left( \Delta_{\mu} \theta_j - A_{j \mu}
\right) \right) \label{zm}
\end{eqnarray}
The Dasgupta-Halperin duality mapping can be easily extended to
this theory. We now write both cosine terms in their Villain
forms, and then proceed as described above. The results
(\ref{zsc1}) and (\ref{zsc2}) continue to have the same form, with
the only change being that the fields $P_{\bar{\jmath}\mu}$ and
$\varphi_{\bar{\jmath}}$ are now also {\em integer\/} valued (and
so must be summed over). Promoting these integer valued fields to
real fields by the Poisson summation method following
Ref.~\cite{jkkn}, we now have to introduce {\em two\/} fugacities:
a vortex fugacity $y_v$ (as before), and a monopole fugacity
$\tilde{y}_m$ (discussed below). Consequently, $\mathcal{Z}_{\rm
SC,3}$ in (\ref{zsc3}) now takes the form
\begin{eqnarray}
 \mathcal{Z}_{M,3} &=& \prod_{\bar{\jmath}} \int_{-\infty}^{\infty} d
b_{\bar{\jmath}\mu} \int_{-\infty}^{\infty} d
\varphi_{\bar{\jmath}} \int_{-\infty}^{\infty} d
\vartheta_{\bar{\jmath}} \exp \left( -\frac{e^2}{2}
\sum_{\bar{\jmath},\mu} \left(b_{\bar{\jmath}\mu} -
\Delta_{\mu} \varphi_{\bar{\jmath}}\right)^2 \right. \nonumber \\
&~&~~~~~~~~~~~~~~~ - \frac{g}{2} \sum_{\Box} \left(
\epsilon_{\mu\nu\lambda} \Delta_{\nu} b_{\bar{\jmath} \lambda}
\right)^2 + y_v \sum_{\bar{\jmath},\mu} \cos \left(2 \pi
b_{\bar{\jmath}\mu} - \Delta_{\mu} \vartheta_{\bar{\jmath}}
\right) \nonumber \\
&~&~~~~~~~~~~~~~~~\left.+ \tilde{y}_m \sum_{\bar{\jmath}} \cos
\left(2 \pi \varphi_{\bar{\jmath}} -
\vartheta_{\bar{\jmath}}\right)\right). \label{zm3}
\end{eqnarray}
Again, the positions of the $\vartheta_{\bar{\jmath}}$ above are
dictated by gauge invariance, and the effect of the vortex and
monopole fugacities is to soften the integer value constraints on
the $b_{\bar{\jmath}\mu}$ and $\varphi_{\bar{\jmath}}$. Proceeding
as described below (\ref{zsc3}), we work in the gauge
$\varphi_{\bar{\jmath}}=0$, and to leading order in $y_v$,
$\tilde{y}_m$ replace $b_{\bar{\jmath}\mu}$ by its saddle point
value in the Gaussian part of the action, which remains
$\overline{b}_{\bar{\jmath}\mu}=0$. Then, instead of (\ref{zsc4}),
we obtain
\begin{equation}
\mathcal{Z}_{M,4} = \prod_{\bar{\jmath}}  \int_{-\infty}^{\infty}
d \vartheta_{\bar{\jmath}} \exp \left( y_v \sum_{\bar{\jmath},\mu}
\cos \left( \Delta_{\mu} \vartheta_{\bar{\jmath}} \right) +
\tilde{y}_m \sum_{\bar{\jmath}} \cos
\left(\vartheta_{\bar{\jmath}}\right)\right). \label{zm4}
\end{equation}
We see that the new second term in (\ref{zm4}) acts like an
ordering field on the dual XY model. Taking the continuum limit as
was done below (\ref{zsc4}) using \cite{hbst} a complex field
$\psi$ conjugate to $e^{i \vartheta_{\bar{\jmath}}}$, now instead
of $\mathcal{Z}_{\rm SC,dual}$ in (\ref{zscd}) we obtain
\cite{nagaosa,sudboxy}
\begin{eqnarray}
\mathcal{Z}_{M,{\rm dual}} &=& \int \mathcal{D}\psi (r, \tau) \exp
\Biggl( - \int d^2 r d \tau \biggl( |\partial_{\mu} \psi|^2 +
\overline{s} |\psi|^2 \nonumber \\
&~&~~~~~~~~~~~~~~~~~~~~~~~~~~~~~~~~~+ \frac{\overline{u}}{2}
|\psi|^4 - y_m (\psi + \psi^{\ast}) \biggr) \Biggr) \label{zmd}
\end{eqnarray}
The new term proportional to $y_m$ has the interpretation of a
{\em monopole fugacity}. The compact gauge field now permits Dirac
monopoles, which are points in spacetime at which vortex loops of
the `superconductor' can end: hence $y_m$ is coupled to the
creation and annihilation operators for the dual boson $\psi$ {\em
i.e.\/} the vortices. In the form (\ref{zmd}) it is also clear
that $y_m$ acts like an ordering field in the dual XY model. We
expect that such an XY model has no phase transition, and $\langle
\psi \rangle \neq 0$ for all $\overline{s}$. So the presence of
monopoles has destroyed the `superconducting' phase. Comparing the
properties of (\ref{zscd}) and (\ref{zmd}) we therefore conclude
that making the gauge field compact in $\mathcal{Z}_{\rm SC}$ in
(\ref{zsc}) is a strongly relevant perturbation: the inverted XY
transition of $\mathcal{Z}_{\rm SC}$ is destroyed in the resulting
model $\mathcal{Z}_{M}$.

\paragraph{C. \underline{Berry phases}}

We are finally ready to face $\mathcal{Z}_1$, and add in the final
layer of complication of the Berry phases. Again, the
Dasgupta-Halperin duality can be extended by combining it with the
methods of Section~\ref{sec:para} (this was partly discussed in
Ref.~\cite{SJ}). Now the monopoles carry Berry phases
\cite{hald88,rs}, and these lead to cancellations among many
monopole configurations. In the long-wavelength limit it turns out
that the only important configurations are those in which the
total monopole magnetic charge is $q$ times the charge of the
elementary monopole \cite{hald88,rs,rsb}. Here $q$ is the smallest
positive integer such that
\begin{equation}
e^{i \pi S q} = 1, \label{qval}
\end{equation}
{\em i.e.\/} $q=4$ for $S$ half an odd integer, $q=2$ for $S$ an
odd integer, and $q=1$ for $S$ an even integer. Using the physical
interpretation of (\ref{zmd}), we therefore conclude that the
monopole fugacity term should be replaced by one in which the
monopoles are created and annihilated in multiples of $q$; the
dual theory of $\mathcal{Z}_1$ in (\ref{z1}) then becomes
\begin{eqnarray}
\mathcal{Z}_{1,{\rm dual}} &=& \int \mathcal{D}\psi (r, \tau) \exp
\Biggl( - \int d^2 r d \tau \biggl( |\partial_{\mu} \psi|^2 +
\overline{s} |\psi|^2 \nonumber \\
&~&~~~~~~~~~~~~~~~~~~~~~~~~~~~~~~~~~+ \frac{\overline{u}}{2}
|\psi|^4 - y_{mq} (\psi^q + \psi^{\ast q}) \biggr) \Biggr)
\label{z1d}
\end{eqnarray}

An explicit derivation of the mapping from $\mathcal{Z}_1$ to
$\mathcal{Z}_{1,{\rm dual}}$ can be obtained by an extension of
the methods described above for $\mathcal{Z}_{\rm SC}$ and
$\mathcal{Z}_{M}$. We express the Berry phase term using the
`background field' $a_{\bar{\jmath}\mu}^0$ in (\ref{d2}), and then
we find that $\mathcal{Z}_{\rm SC,2}$ in (\ref{zsc2}) is now
replaced by
\begin{eqnarray}
&& \mathcal{Z}_{1,2} =  \sum_{\{b_{\bar{\jmath}\mu}\}}
\sum_{\{\varphi_{\bar{\jmath}}\}} \exp \left( -\frac{e^2}{2}
\sum_{\bar{\jmath},\mu} \left(b_{\bar{\jmath}\mu} -
\Delta_{\mu} \varphi_{\bar{\jmath}} - 2S a_{\bar{\jmath}\mu}^0 \right)^2 \right. \nonumber \\
&&~~~~~~~~~~~~~~~~~~~~~~~~~~~~~~~~~~~~~~~~~~\left. - \frac{g}{2}
\sum_{\Box} \left( \epsilon_{\mu\nu\lambda} \Delta_{\nu}
b_{\bar{\jmath} \lambda} \right)^2 \right). \label{z12}
\end{eqnarray}
Notice that, as in Section~\ref{sec:para}, the Berry phases appear
as offsets in the dual action. We now promote the integer field
$b_{\bar{\jmath}\mu}$ and $\varphi_{\bar{\jmath}}$ to real fields
by the Poisson summation method (just as in (\ref{zm3})), at the
cost of introducing vortex and monopole fugacities. The final
steps, following the procedure below (\ref{zm3}), are to transform
to the gauge $\varphi_{\bar{\jmath}}=0$, and to then set the
`Higgsed' dual gauge field $b_{\bar{\jmath}\mu}$ to its saddle
point value determined from the Gaussian terms in the action. It
is the latter step which is now different, and the presence of the
$a_{\bar{\jmath}\mu}^0$ now implies that the saddle point value
$\overline{b}_{\bar{\jmath}\mu}$ will be non-zero and site
dependent. Indeed, it is crucial that the saddle point be
determined with great care, and that the square lattice symmetry
of the underlying problem be fully respected. This saddle point
determination is in many ways analogous to the computation in
Section III.B of Ref.~\cite{rsb}, and it is important that all the
modes on the lattice scale be fully identified in a similar
manner. The similarity to Ref.~\cite{rsb} becomes clear after
using the parameterization in (\ref{XY}) for
$a_{\bar{\jmath}\mu}^0$ in terms of the
$\mathcal{X}_{\bar{\jmath}}$ and the $\mathcal{Y}_{j\mu}$ shown in
Fig~\ref{fig9}. Finally, after transforming $b_{\bar{\jmath}\mu}
\rightarrow b_{\bar{\jmath}\mu} + 2 S \Delta_{\mu}
\mathcal{X}_{\bar{\jmath}}$ and $\vartheta_{\bar{\jmath}}
\rightarrow \vartheta_{\bar{\jmath}} + 4 \pi S
\mathcal{X}_{\bar{\jmath}}$, we obtain from (\ref{z12})
\begin{eqnarray}
 \mathcal{Z}_{1,3} &=& \prod_{\bar{\jmath}} \int_{-\infty}^{\infty} d
b_{\bar{\jmath}\mu}  \int_{-\infty}^{\infty} d
\vartheta_{\bar{\jmath}} \exp \left( -\frac{e^2}{2}
\sum_{\bar{\jmath},\mu} \left(b_{\bar{\jmath}\mu} -
2 S \epsilon_{\mu\nu\lambda} \Delta_{\nu} \mathcal{Y}_{\lambda} \right)^2 \right. \nonumber \\
&~&~~~~~~~~~~~~~~~ - \frac{g}{2} \sum_{\Box} \left(
\epsilon_{\mu\nu\lambda} \Delta_{\nu} b_{\bar{\jmath} \lambda}
\right)^2 + y_v \sum_{\bar{\jmath},\mu} \cos \left(2 \pi
b_{\bar{\jmath}\mu} - \Delta_{\mu} \vartheta_{\bar{\jmath}}
\right) \nonumber \\
&~&~~~~~~~~~~~~~~~\left.+ \tilde{y}_m \sum_{\bar{\jmath}} \cos
\left(\vartheta_{\bar{\jmath}} + 4 \pi S
\mathcal{X}_{\bar{\jmath}}\right)\right). \label{z13}
\end{eqnarray}
Now, the saddle point value of the massive field
$b_{\bar{\jmath}\mu}$ is easily determined from the first terms in
(\ref{z13}), yielding
\begin{equation}
\overline{b}_{\bar{\jmath}\mu} = \alpha \epsilon_{\mu\nu\lambda}
\Delta_{\nu}\mathcal{Y}_{j \lambda}. \label{saddleb}
\end{equation}
where $\alpha \equiv 2S e^2 / (e^2 + 8g)$. Note that only the
spatial components of $\overline{b}_{\bar{\jmath}\mu}$ are
non-zero, and these have the simple structure of Fig~\ref{fig9}c.
In particular, the magnitude of the
$\overline{b}_{\bar{\jmath}\mu}$ are the same on all the spatial
links, and the use of (\ref{XY}) was crucial in obtaining this
appealing result. With this saddle point value, (\ref{z13})
simplifies to the following model for the field
$\vartheta_{\bar{\jmath}}$ only (this is the form of (\ref{zm4})
after accounting for Berry phases):
\begin{eqnarray}
\mathcal{Z}_{1,4} &=& \prod_{\bar{\jmath}}
\int_{-\infty}^{\infty} d \vartheta_{\bar{\jmath}} \exp \left( y_v
\sum_{\bar{\jmath},\mu} \cos \left( \Delta_{\mu}
\vartheta_{\bar{\jmath}} - 2 \pi
\overline{b}_{\bar{\jmath}\mu}\right) \right. \nonumber \\
&~&~~~~~~~~~~~~~~~~~~~~~~~~~~~~~\left. + \tilde{y}_m
\sum_{\bar{\jmath}} \cos \left(\vartheta_{\bar{\jmath}}+ 4 \pi S
\mathcal{X}_{\bar{\jmath}}\right)\right). \label{z14}
\end{eqnarray}
The most important property of this dual XY model is the nature of
the ordering field in the last term of (\ref{z14}). For $S=1/2$,
notice from Fig~\ref{fig9}a that this field is oriented
north/east/south/west on the four sublattices in of the dual
lattice in Fig~\ref{fig9}c. So if we take a naive continuum limit,
the average field vanishes! This is the key effect responsible for
the cancellations among monopole configurations induced by Berry
phases noted earlier; in the dual formulation, the Berry phases
have appeared in differing orientations of the dual ordering
field. The XY model in (\ref{z14}) also has the contribution from
$\overline{b}_{\bar{\jmath}\mu}$, which appear as a `staggered
flux' acting on the $\vartheta_{\bar{\jmath}}$ (see
Fig~\ref{fig9}c), but we now show that this is not as crucial in
the continuum limit.

Before we take the continuum limit of $\mathcal{Z}_{1,4}$, we
discuss its implementation of the square lattice symmetries. In
particular, we are interested in the $Z_4$ symmetry which rotates
the four sublattices in Fig~\ref{fig9}c into each other, as the
values of $\mathcal{X}_{\bar{\jmath}}$ seem to distinguish between
them. Let us consider the symmetry $\mathcal{R}_n$ which rotates
lattice anticlockwise by an angle $n \pi /2$ about the direct
lattice point at the center of a plaquette in Fig~\ref{fig9}c,
associated with the transformation in (\ref{hf1}). It is easy to
see that $\mathcal{Z}_{1,4}$ remains invariant under
$\mathcal{R}_n$ provided we simultaneously rotate the angular
variables $\vartheta_{\bar{\jmath}}$:
\begin{equation}
\mathcal{R}_n : ~~~~\vartheta_{\bar{\jmath}} \rightarrow
\vartheta_{\bar{\jmath}} + n S \pi \label{rnt}
\end{equation}
It is now useful to introduce complex variables which realize
irreducible representations of this $Z_4$ symmetry. We divide the
lattice into plaquettes like those in Fig~\ref{fig9}c, and for
each plaquette we define variables $\psi_p$, with $p$ integer, by
\begin{equation}
\psi_p = \frac{1}{2} \left( e^{i \vartheta_W} + e^{ip\pi/2} e^{i
\vartheta_X} + e^{ip\pi} e^{i \vartheta_Y} + e^{i3p\pi/2} e^{i
\vartheta_Z} \right). \label{defpsip}
\end{equation}
Note that we need only use $p=0,1,2,3$ because $\psi_p$ depends
only on $p \mbox{(mod 4)}$. Under the symmetry $\mathcal{R}_n$ we
clearly have
\begin{equation}
\mathcal{R}_n: ~~~\psi_p \rightarrow e^{in(2S-p)\pi/2} \psi_p;
\label{rn2}
\end{equation}
the factor of $e^{i n S \pi}$ arises from (\ref{rnt}), and that of
$e^{-in p\pi/2}$ from the real-space rotation of the lattice
points. Note that only for $p = 2S$ is $\psi_p$ invariant under
$\mathcal{R}_n$,  and this is consistent with the fact that it is
$\psi_{2S}$ which appears in $\mathcal{Z}_{1,4}$ as the ordering
field term. Let us now write the action in $\mathcal{Z}_{1,4}$ in
terms of these new variables. Ignoring the spacetime variation
from plaquette to plaquette, the action per plaquette is
\begin{equation}
\mathcal{S}_{1,4} = -2 y_v \sum_{p=0}^3 \left[ \cos\left(\pi(
p-\alpha)/2\right) \left| \psi_p \right|^2 \right]- \tilde{y}_m
\left( \psi_{2S} + \psi_{2S}^{\ast} \right) + \ldots \label{s14}
\end{equation}
Here the ellipses represent other allowed terms, all consistent
with the symmetry (\ref{rn2}), which must be included to implement
the (softened) constraints on $\psi_p$ arising from
(\ref{defpsip}) and the fact that the $e^{i
\vartheta_{\bar{\jmath}}}$ are unimodular. Apart from $\psi_{2S}$,
for which there is already an ordering field in the action, the
condensation of any of the other $\psi_p$ breaks the lattice
symmetry (\ref{rn2}), and so drives a quantum phase transition to
the bond-ordered state. The choice among the $\psi_p$ is
controlled by the co-efficient of the $y_v$ term in (\ref{s14}),
and we choose the value of $p\neq 2S$ for which
$\cos\left(\pi(\alpha + p)/2\right)$ is a maximum. We are
interested in the large $g$ paramagnetic phase, and here $\alpha$
is small, and the appropriate value is $p=0$. The resulting
continuum theory for $\psi = \psi_0$ then must be invariant under
(\ref{rn2}), and it is easily seen that this has just the form
$\mathcal{Z}_{1,{\rm dual}}$ in (\ref{z1d}) with $q$ determined by
(\ref{qval}). Other choices of $p$ for the order parameter lead to
different types of bond order, with a ground state degeneracy
smaller or larger than the $q$ in (\ref{qval}); such states have
partial or additional bond order, and are clearly possible in
general. However, our analysis of the paramagnetic states in
Section~\ref{sec:para} indicates that a choice $\psi = \psi_{p
\neq 0}$ is unlikely for the models under consideration here, and
we will not consider this case further here.

\begin{table}[t]
\begin{tabular}{p{2.25in}|p{2.25in}}
\hline
\multicolumn{2}{c}{\rule[-3mm]{0mm}{8mm} $N=1$} \\
\hline \multicolumn{1}{c|}{\rule[-3mm]{0mm}{8mm}\bf Direct lattice
model} &
\multicolumn{1}{c}{\rule[-3mm]{0mm}{8mm}\bf Dual model} \\
\hline  \begin{eqnarray*} \mathcal{L}_{SC} &=& (1/(2e^2)) \left(
\epsilon_{\mu\nu\lambda}\Delta_{\nu} A_{j \lambda} \right)^2
\\&-& (1/g) \cos \left( \Delta_{\mu} \theta_j - A_{j \mu} \right)
\end{eqnarray*} & \begin{eqnarray*} \mathcal{L}_{SC,{\rm dual}} = |\partial_{\mu} \psi|^2 +
\overline{s} |\psi|^2 + \frac{\overline{u}}{2} |\psi|^4 \end{eqnarray*} \\
\hline \begin{eqnarray*} \mathcal{L}_{M} &=& -(1/e^2) \cos \left(
\epsilon_{\mu\nu\lambda}\Delta_{\nu} A_{j \lambda} \right)
\\&-& (1/g) \cos \left( \Delta_{\mu} \theta_j - A_{j \mu} \right)
\end{eqnarray*} & \begin{eqnarray*} \mathcal{L}_{M,{\rm dual}} &=& |\partial_{\mu} \psi|^2 +
\overline{s} |\psi|^2 + \frac{\overline{u}}{2} |\psi|^4 \nonumber
\\ &-& y_m (\psi + \psi^{\ast} )
\end{eqnarray*} \\
\hline
\begin{eqnarray*} \mathcal{L}_{1} &=& -(1/e^2) \cos \left(
\epsilon_{\mu\nu\lambda}\Delta_{\nu} A_{j \lambda} \right)
\\&-& (1/g) \cos \left( \Delta_{\mu} \theta_j - A_{j \mu} \right)
- i 2S \eta_j A_{j \tau}
\end{eqnarray*} & \begin{eqnarray*} \mathcal{L}_{1,{\rm dual}} &=& |\partial_{\mu} \psi|^2 +
\overline{s} |\psi|^2 + \frac{\overline{u}}{2} |\psi|^4 \nonumber
\\ &-& y_{mq} (\psi^q + \psi^{\ast q} )
\end{eqnarray*} \\
\hline
\end{tabular}
 \caption{Summary of the duality mappings for $N=1$. Only the Lagrangean's
are specified, and a summation/integration of these over
spacetimes is implicit. The fixed field $\eta_j=\pm 1$ in the
Berry phase in the third row is the sublattice staggering factor
in (\protect\ref{neel}). The integer $q$ in the third row is
specified in (\protect\ref{qval}). For $S=1/2$, we have $q=4$, and
then the $y_{mq}$ perturbation is dangerously irrelevant. Hence
the critical theory for the model with monopoles and Berry phases
in the third row, is identical to that for the first row.}
\end{table}
We have now completed our promised derivation of the model
$\mathcal{Z}_{1,{\rm dual}}$ in (\ref{z1d}) dual to the $N=1$
lattice gauge theory model $\mathcal{Z}_1$ in (\ref{z1}). Rather
than being an XY model in a field (as in (\ref{zmd})),
$\mathcal{Z}_{1,{\rm dual}}$ is an XY model with a $q$-fold
anisotropy. This anisotropy encapsulates the $q$-fold binding of
monopoles claimed earlier. In the language of (\ref{z14}) the
average ordering fields on the $\vartheta_{\bar{\jmath}}$
oscillate from site to site and cancel out, and only the $q$-th
moment of the field survives. Now the combined effect of the
monopoles and Berry phases in $\mathcal{Z}_1$ is decided by the
term proportional to $y_{mq}$. In the paramagnetic phase of the
direct model, which is $\overline{s} < \overline{s}_c$ and
$\langle \psi \rangle \neq 0$, this $q$-fold anisotropy is
certainly very important. For $S=1/2$, $q=4$ it orders the $\psi$
field along four particular angles, and these are easily shown to
be \cite{SJ} one of the four degenerate bond-ordered states in
Fig~\ref{fig11}. However, at the critical point
$\overline{s}=\overline{s}_c$ it is known that this 4-fold
anisotropy is irrelevant \cite{vicari}: so in $\mathcal{Z}_{1,{\rm
dual}}$ the monopoles can be neglected at the critical point
$s=s_c$, but not away from it.

We have now achieved the desired objective of this subsection.
Compactness alone was a strongly relevant perturbation on the
model of a scalar field coupled to electromagnetism in
$\mathcal{Z}_{\rm SC}$. However, when we combined compactness with
the Berry phases in $\mathcal{Z}_{1}$, then we found that the
monopoles effectively cancelled each other out at the critical
point for $S=1/2$. Consequently the theory for the critical point
in $\mathcal{Z}_1$ is identical to the theory for the critical
point in $\mathcal{Z}_{\rm SC}$, and this is the simple inverted
XY model $\mathcal{Z}_{\rm SC,dual}$ in (\ref{zscd}). The results
of this subsection are summarized in Table 1.

\subsubsection{4.2.2~Easy plane model at $N=2$}
\label{n2}

A second explicit example of the remarkable phenomenon described
above is provided by the physically relevant $N=2$ case of the
model of central interest, $\mathcal{Z}$ in (\ref{z}), but in the
presence of an additional spin-anisotropy term preferring that the
spins lie within the XY plane. In such a situation, we may write
the complex spinor $z_{j a}$ as
\begin{equation}
z_{j a} = \frac{1}{\sqrt{2}} \left( \begin{array}{c} e^{i
\theta_{j \uparrow}}
\\ e^{i \theta_{j \downarrow}} \end{array} \right),
\label{zt12}
\end{equation}
so that the action is expressed in terms of {\em two} angular
fields, $\theta_{\uparrow}$ and $\theta_{\downarrow}$. Inserting
(\ref{zt12}) in (\ref{z}), we obtain a generalization of the $N=1$
model $\mathcal{Z}_1$ in (\ref{z1}):
\begin{eqnarray}
\mathcal{Z}_2 &=& \prod_j \int_0^{2 \pi} \frac{d\theta_{j \uparrow
}}{2 \pi} \int_0^{2 \pi} \frac{d\theta_{j \downarrow }}{2 \pi}
\int_0^{2 \pi} \frac{dA_{j \mu}}{2 \pi} \exp \left( \frac{1}{e^2}
\sum_{\Box} \cos \left( \epsilon_{\mu\nu\lambda}\Delta_{\nu} A_{j
\lambda}
\right) \right. \nonumber \\
&~&~~~~~~~~~~~~~~ \left.+ \frac{1}{2g} \sum_{j, \mu,a} \cos \left(
\Delta_{\mu} \theta_{j a} - A_{j \mu} \right) + i 2S \sum_j \eta_j
A_{j \tau} \right). \label{z2}
\end{eqnarray}
As in (\ref{z1}), we have chosen to explicitly include a Maxwell
term for the U(1) gauge field as it proves convenient in the
subsequent duality analysis. The model $\mathcal{Z}_2$ provides a
complete description of the phases of the square lattice
antiferromagnet (\ref{hams}) with an additional easy-plane
anisotropy term.

We can now proceed with a duality analysis of (\ref{z2}) using
methods precisely analogous to those discussed in
Section~\ref{n1}.1: the only difference is we now have two angular
fields $\theta_{a=\uparrow,\downarrow}$, and so certain fields
come with two copies. We will therefore not present any details,
and simply state the series of results which appear here, which
closely parallel those obtained above for $N=1$.
\begin{itemize}
\item
Neglecting {\em both} compactness of the $U(1)$ gauge field
and the Berry phases, it is straightforward to take the continuum
limit of $\mathcal{Z}_2$ in its direct representation, and we
obtain the theory $\mathcal{Z}_c$ in (\ref{zc}), but with an
additional spin-anisotropy term
\begin{eqnarray}
\mathcal{Z}_{2c} &=& \int \mathcal{D} z_a (r, \tau) \mathcal{D}
A_{\mu} (r, \tau) \exp \Biggl( - \int d^2 r d \tau \biggl[
|(\partial_\mu -
i A_{\mu}) z_a |^2 + s |z_a |^2  \nonumber \\
&~&~~~~~~~~~~~~~~~+ \frac{u}{2} (|z_a|^2)^2 + v |z_\uparrow |^2
|z_\downarrow |^2 +  \frac{1}{4e^2} (\epsilon_{\mu\nu\lambda}
\partial_\nu A_\lambda )^2 \biggl]\Biggl), \label{z2c}
\end{eqnarray}
where $v > 0$ prefers spins in the easy plane. We can carry
through the analog of the duality mapping between (\ref{zsc}) and
(\ref{zscd}), and instead of (\ref{zscd}) we now obtain a theory
for {\em two} dual fields $\psi_a$ representing vortices in
$\theta_\uparrow$ and $\theta_\downarrow$ \cite{mv}
\begin{eqnarray}
&& \mathcal{Z}_{2c,{\rm dual}} = \int \mathcal{D} \psi_a (r, \tau)
\mathcal{D} B_{\mu} (r, \tau) \exp \Biggl( - \int d^2 r d \tau
\biggl[ |(\partial_\mu - i B_{\mu}) \psi_\uparrow |^2 \nonumber \\
&~&~~~~~~~~~~~~~~+ |(\partial_\mu + i B_{\mu}) \psi_\downarrow
|^2+ \overline{s} |\psi_a |^2  + \frac{\overline{u}}{2}
(|\psi_a|^2)^2 + \overline{v} |\psi_\uparrow |^2 |\psi_\downarrow
|^2 \nonumber
\\&~&~~~~~~~~~~~~~~~~~~~~~~~~~~~~~~~~~~~~~~~+ \frac{1}{4\overline{e}^2} (\epsilon_{\mu\nu\lambda}
\partial_\nu B_\lambda )^2 \biggl]\Biggl). \label{z2cd}
\end{eqnarray}
Note that there is now a non-compact U(1) gauge field $B_\mu$
which survives the continuum limit: this field arises from the
analog of the field $b_{\bar{\jmath} \mu}$ in (\ref{zsc3}), and
here it is not completely Higgsed out. The most remarkable
property of (\ref{z2cd}) is that it is identical in structure to
(\ref{z2c}): the actions are identical under the mapping
$z_\uparrow \rightarrow \psi_\uparrow$, $z_\downarrow \rightarrow
\psi_\downarrow^{\ast}$, and $A_\mu \rightarrow B_\mu$. In other
words, the theory $\mathcal{Z}_{2c}$ is {\em self-dual} \cite{mv}.
\item
As in Section~\ref{n1}.1, we next make the $A_{\mu}$ gauge
field compact, but continue to ignore Berry phases {\em i.e.} we
perform a duality analysis on (\ref{z2}), in the absence of the
last term in the action. Now, instead of (\ref{zmd}), (\ref{z2cd})
is modified to
\begin{eqnarray}
&& \mathcal{Z}_{2M,{\rm dual}} = \int \mathcal{D} \psi_a (r, \tau)
\mathcal{D} B_{\mu} (r, \tau) \exp \Biggl( - \int d^2 r d \tau
\biggl[ |(\partial_\mu - i B_{\mu}) \psi_\uparrow |^2 \nonumber \\
&~&~~~~~~~~~~~~~~+|(\partial_\mu + i B_{\mu}) \psi_\downarrow |^2
+\overline{s} |\psi_a |^2 + \frac{\overline{u}}{2} (|\psi_a|^2)^2
+
\overline{v} |\psi_\uparrow |^2 |\psi_\downarrow |^2  \nonumber \\
&~&~~~~~~~~~~~~~~+ \frac{1}{4\overline{e}^2}
(\epsilon_{\mu\nu\lambda}
\partial_\nu B_\lambda )^2 - y_m ( \psi_{\uparrow}
\psi_\downarrow + \psi_{\downarrow}^{\ast} \psi_\uparrow^\ast)
\biggl]\Biggl). \label{z2md}
\end{eqnarray}
The last term represents the influence of monopoles, and these now
have the effect of turning a $\psi_\uparrow$ vortex into a
$\psi_\downarrow$ vortex \cite{mv,senthil,glw}. Again, as in
(\ref{zmd}), the $y_m$ term in (\ref{z2md}) is clearly a strongly
relevant perturbation to $\mathcal{Z}_{2c,{\rm dual}}$ in
(\ref{z2cd}). It ties the phases of $\psi_\uparrow$ and
$\psi_\downarrow$ to each other, so that (\ref{z2md}) is
effectively the theory of a {\em single} complex scalar coupled to
a non-compact U(1) gauge field $B_{\mu}$. However, we have already
considered such a theory in the {\em direct} representation in
(\ref{zsc}). We can now move from the dual representation in
(\ref{z2md}) back to the direct representation, by the mapping
between (\ref{zsc}) and (\ref{zscd}). This leads to the
conclusion, finally, that the theory (\ref{z2md}) is dual to an
ordinary XY model. In other words, the theory $\mathcal{Z}_2$ in
(\ref{z2}) without its Berry phase term is an XY model. However,
this is precisely the expected conclusion, and could have been
easily reached without this elaborate series of duality mappings:
just integrating over $A_{j \mu}$ for large $e^2$ yields an XY
model in the angular field $\theta_\uparrow - \theta_\downarrow$,
which represents the orientation of the physical in-plane N\'{e}el
order.
\item
Finally, let us look at the complete theory
$\mathcal{Z}_2$. An explicit duality mapping can be carried out,
and as in (\ref{z1d}), the action (\ref{z2md}) is replaced by
\cite{lfs,sp,senthil,glw}
\begin{eqnarray}
&& \mathcal{Z}_{2M,{\rm dual}} = \int \mathcal{D} \psi_a (r, \tau)
\mathcal{D} B_{\mu} (r, \tau) \exp \Biggl( - \int d^2 r d \tau
\biggl[ |(\partial_\mu - i B_{\mu}) \psi_\uparrow |^2 \nonumber \\
&~&~~~~~~~~~~~~~~+|(\partial_\mu + i B_{\mu}) \psi_\downarrow |^2
+\overline{s} |\psi_a |^2 + \frac{\overline{u}}{2} (|\psi_a|^2)^2
+
\overline{v} |\psi_\uparrow |^2 |\psi_\downarrow |^2  \nonumber \\
&~&~~~~~~~~~~~~~~+ \frac{1}{4\overline{e}^2}
(\epsilon_{\mu\nu\lambda}
\partial_\nu B_\lambda )^2 - y_{mq} \left( \left(\psi_{\uparrow}
\psi_\downarrow \right)^q + \left(\psi_{\downarrow}^{\ast}
\psi_\uparrow^\ast \right)^q \right) \biggl]\Biggl), \label{z2d}
\end{eqnarray}
where the integer $q$ was defined in (\ref{qval}). The subsequent
reasoning is the precise analog of that for $N=1$. For $S=1/2$ and
$q=4$, the term proportional to $y_{mq}$ representing $q$-fold
monopole is irrelevant at the critical point (but not away from it
in the paramagnetic phase). Consequently, the critical theory of
(\ref{z2d}) reduces to (\ref{z2cd}). So just as at $N=1$, the {\em
combined} influence of monopoles and Berry phases is dangerously
irrelevant at the critical point, and for the critical theory we
can take a naive continuum limit of $\mathcal{Z}_2$ neglecting
both the Berry phases and the compactness of the gauge field.
\end{itemize}
We have now completed our discussion of the $N=2$ easy plane model
and established the existence of the same remarkable phenomenon
found in Section~\ref{n1}.1 for $N=1$, and claimed more generally
\cite{senthil,glw} at the beginning of Section~\ref{sec:qpt} as
the justification for the critical theory (\ref{zc}). As we saw in
some detail in Section~\ref{sec:para}, monopoles, and attendant
Berry phases, are absolutely crucial in understanding the onset of
confinement and bond order in the paramagnetic phase. However, for
$S=1/2$, the Berry phases induce a destructive quantum
interference between the monopoles at the quantum critical point,
leading to a critical theory with `deconfined' spinons and a
non-compact U(1) gauge field which does not allow monopoles. These
results are summarized in Table 2.
\begin{table}[t]
\begin{tabular}{p{2.25in}|p{2.25in}}
\hline \multicolumn{2}{c}{\rule[-3mm]{0mm}{8mm} $N=2$, easy plane} \\
\hline \multicolumn{1}{c|}{\rule[-3mm]{0mm}{8mm}\bf Direct lattice
model} &
\multicolumn{1}{c}{\rule[-3mm]{0mm}{8mm}\bf Dual model} \\
\hline \begin{eqnarray*} \mathcal{L}_{2,SC} &=& (1/(2e^2)) \left(
\epsilon_{\mu\nu\lambda}\Delta_{\nu} A_{j \lambda} \right)^2
\\&-& (1/g) \cos \left( \Delta_{\mu} \theta_{ja} - A_{j \mu} \right)
\end{eqnarray*} & \begin{eqnarray*} && \mathcal{L}_{2SC,{\rm dual}} =
|(\partial_{\mu}-i B_\mu) \psi_\uparrow|^2 \nonumber \\ && +
|(\partial_{\mu}+i B_\mu) \psi_\downarrow|^2 + \overline{s}
|\psi_a|^2 + \frac{\overline{u}}{2}
\left(|\psi_a|^2 \right)^2 \nonumber \\ &&
+ \overline{v} |\psi_\uparrow |^2 |\psi_\downarrow |^2  + \frac{1}{2 \overline{e}^2}
( \epsilon_{\mu\nu\lambda} \partial_\nu B_\lambda )^2 \end{eqnarray*} \\
\hline \begin{eqnarray*} \mathcal{L}_{2M} &=& -(1/e^2) \cos \left(
\epsilon_{\mu\nu\lambda}\Delta_{\nu} A_{j \lambda} \right)
\\&-& (1/(2g)) \cos \left( \Delta_{\mu} \theta_{ja} - A_{j \mu} \right)
\end{eqnarray*} & \begin{eqnarray*} && \mathcal{L}_{2M,{\rm dual}} = |(\partial_{\mu}-i B_\mu) \psi_\uparrow|^2 \nonumber \\ && +
|(\partial_{\mu}+i B_\mu) \psi_\downarrow|^2 + \overline{s}
|\psi_a|^2 + \frac{\overline{u}}{2} \left(|\psi_a|^2 \right)^2
\nonumber \\ && + \overline{v} |\psi_\uparrow |^2 |\psi_\downarrow
|^2  + \frac{1}{2 \overline{e}^2} ( \epsilon_{\mu\nu\lambda}
\partial_\nu B_\lambda )^2
\\ &&~~~- y_m \left( \psi_\uparrow \psi_\downarrow + \psi_\uparrow^{\ast} \psi_\downarrow^\ast \right)
\end{eqnarray*} \\
\hline
\begin{eqnarray*} && \mathcal{L}_{2} = -(1/e^2) \cos \left(
\epsilon_{\mu\nu\lambda}\Delta_{\nu} A_{j \lambda} \right)
\\&&- (1/g) \cos \left( \Delta_{\mu} \theta_{ja} - A_{j \mu} \right)
- i 2S \eta_j A_{j \tau}
\end{eqnarray*} & \begin{eqnarray*} && \mathcal{L}_{2,{\rm dual}} = |(\partial_{\mu}-i B_\mu) \psi_\uparrow|^2 \nonumber \\ && +
|(\partial_{\mu}+i B_\mu) \psi_\downarrow|^2 + \overline{s}
|\psi_a|^2 + \frac{\overline{u}}{2} \left(|\psi_a|^2 \right)^2
\nonumber \\ && + \overline{v} |\psi_\uparrow |^2 |\psi_\downarrow
|^2  + \frac{1}{2 \overline{e}^2} ( \epsilon_{\mu\nu\lambda}
\partial_\nu B_\lambda )^2
\\ &&~~~- y_{mq} \left( \left(\psi_\uparrow \psi_\downarrow \right)^q +
\left( \psi_\uparrow^{\ast} \psi_\downarrow^\ast \right)^q \right)
\end{eqnarray*} \\
\hline
\end{tabular}
\caption{As in Table 1, but for the $N=2$ easy plane case. The
index $a$ extends over the two values $\uparrow$, $\downarrow$.
Again for $S=1/2$, $q=4$, the critical theory for the third row is
the same as that for the first row. The dual model in the second
row is effectively the theory of a single complex scalar coupled
to a non-compact U(1) gauge field $B_\mu$; by the inverse of the
duality mapping in the first row of Table 1, this theory has a
direct XY transition.}
\end{table}

The results in Tables 1 and 2 can be generalized to arbitrary
values of $N$, for models with the analog of an `easy plane'
anisotropy: as in (\ref{zt12}), all the $z_a$ have equal modulus
and are expressed in terms of $a =1 \ldots N$ angles $\theta_a$.
The dual models have $N$ vortex fields $\psi_a$, and $N-1$
non-compact U(1) gauge fields $B_{b \mu}$, $b=1 \ldots (N-1)$. For
$a=1\ldots (N-1)$, the field $\psi_a$ has a charge $+1$ under the
gauge field with $b=a$, and is neutral under all gauge fields with
$b\neq a$. For $a=N$, the field $\psi_N$, has a charge $-1$ under
{\em all} $N-1$ gauge fields. (This gauge structure is similar to
that found in `moose' field theories \cite{nima}.) The dual
representation of the monopole operator is $\prod_{a=1}^N \psi_a$,
and this appears as the co-efficient of $y_m$ (notice that this
operator is neutral under all the gauge fields). The $q^{\rm th}$
power of this operator appears as the coefficient of $y_{mq}$.
Note that the monopole operators involves a product of $N$ fields,
and for large enough $N$, {\em both} $y_m$ and $y_{mq}$ can be
expected to be irrelevant perturbations at the quantum critical
point.

Finally, these analyses of $\mathcal{Z}$ in (\ref{z}) can be
complemented by a study of its $N\rightarrow \infty$ limit,
without any easy-plane anisotropy. This was carried out some time
ago \cite{MS,SJ}, and it was found that monopoles were dangerously
irrelevant at the quantum critical point, {\em both} with or
without Berry phases (as noted above for large $N$ in the easy
plane case). It is important to note that the situation at large
$N$ is subtly different from that for $N=1,2$: in the latter case,
monopoles are dangerously irrelevant in the presence of $S=1/2$
Berry phases, but relevant without Berry phases. The key
understanding of this distinction emerged in the recent work of
Senthil {\em et al.} \cite{senthil,glw}, which finally succeeded
in placing the earlier large $N$ results within the context of
dual theories of topological defects in statistical mechanics.

\section{Triangular lattice antiferromagnet}
\label{sec:triangle}

We continue our analysis of quantum antiferromagnets with an odd
number of $S=1/2$ spins per unit cell, but consider a class
qualitatively different from those in Section~\ref{sec:square}.
One of the defining properties of the models of
Section~\ref{sec:square} was that the magnetically ordered
N\'{e}el state was defined by (\ref{neel}): the average magnetic
moment on all sites were collinear, and only a single vector ${\bf
n}$ was required to specify the orientation of the ground state.
This section shall consider models in which the moments are {\em
non-collinear}; the triangular lattice is the canonical example.
However, similar results should also apply to other
two-dimensional lattices with non-collinear ground states, such as
the distorted triangular lattice found in Cs$_2$CuCl$_4$
\cite{radu}.

\begin{figure}[t]
\centerline{\includegraphics[width=4.5in]{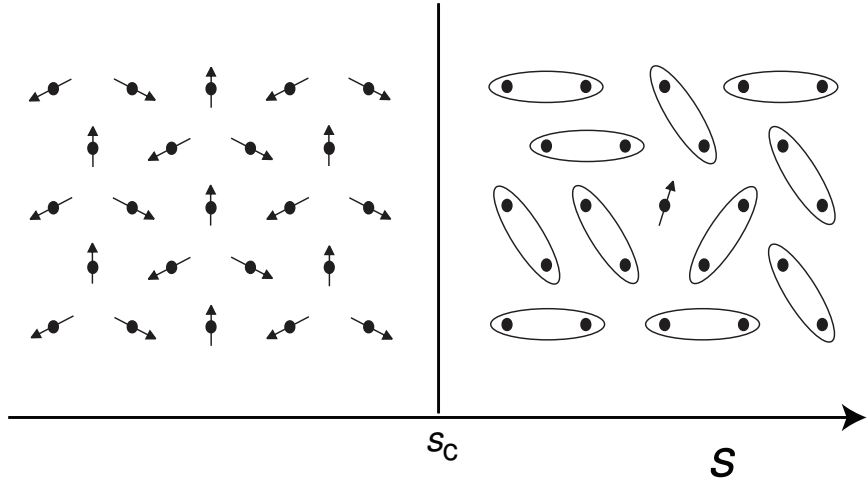}}
\caption{Quantum phase transition described by $\mathcal{Z}_w$ in
(\protect\ref{zw}) as a function of $s$. The state on the left has
non-collinear magnetic order described by (\protect\ref{neelt}).
The state on the right is a `resonating valence bond' (RVB)
paramagnet with topological order and fractionalized $S=1/2$
neutral spinon excitations (one spinon is shown above). Such a
magnetically ordered state is observed in Cs$_2$CuCl$_3$
\protect\cite{radu,radu2}, and there is evidence that the higher
energy spectrum can be characterized in terms of excitations of
the RVB state \protect\cite{ybkim}.} \label{fig13}
\end{figure}

We consider the model (\ref{hams}), but with the spins residing on
the sites of the triangular lattice. This has a magnetically
ordered state illustrated in Fig~\ref{fig13}; for this state
(\ref{neel}) is replaced by
\begin{equation}
\left \langle {\bf S}_j \right \rangle = N_0 \left( {\bf n}_1 \cos
(\vec Q \cdot \vec r) + {\bf n}_2 \sin ( \vec Q \cdot \vec r )
\right). \label{neelt}
\end{equation}
Here $\vec Q = 2 \pi (1/3, 1/\sqrt{3})$ is the ordering
wavevector, and ${\bf n}_{1,2}$ are two arbitrary orthogonal unit
vectors in spin space
\begin{equation}
{\bf n}_1^2 = {\bf n}_2^2 = 1~~~~;~~~~{\bf n}_1 \cdot {\bf n}_2 =
0. \label{constn}
\end{equation}
A distinct ground state, breaking spin rotation symmetry, is
obtained for each choice of ${\bf n}_{1,2}$.

We now wish to allow the values of ${\bf n}_{1,2}$ to fluctuate
quantum mechanically across spacetime, ultimately producing a
paramagnetic state. As in Section~\ref{sec:square}, we should
account for the Berry phases of each spin while setting up the
effective action: an approach for doing this is presented in
Section VI of Ref.~\cite{sp}. However, the full structure of the
critical theory is not understood in all cases, as we describe
below.

One possible structure of the paramagnetic state is a confining,
bond-ordered state, similar to that found in
Section~\ref{sec:square}. However, there is no complete theory for
a possible direct second-order transition from a non-collinear
magnetically ordered state to such a paramagnet. Ignoring Berry
phases, one could define the complex field $\Phi_{\alpha} =
n_{1\alpha} + i n_{2 \alpha}$, which, by (\ref{constn}), obeys
$\Phi_{\alpha}^2 = 0$, and then proceed to write down an effective
action with the structure of (\ref{sp2}). However, it is clear
that such a theory describes a transition to a paramagnetic phase
with a doublet of $S=1$ triplet quasiparticles, and we can
reasonably expect that such a phase has spontaneous bond order (in
contrast to the explicit dimerization in the models of
Section~\ref{sec:dimer}). Berry phases surely play an important
role in inducing this bond order (as they did in
Section~\ref{sec:para}), but there is no available theory for how
this happens in the context of (\ref{sp2}). Indeed, it is possible
that there is no such direct transition between the non-collinear
antiferromagnet and the bond-ordered paramagnet, and resolving
this issue remains an important open question.

In contrast, it is possible to write down a simple theory for a
direct transition between the non-collinear antiferromagnet and a
paramagnetic phase not discussed so far: a resonating valence bond
liquid \cite{pwa,krs,sondhi} with {\em deconfined spinons and
topological order}. This theory is obtained by observing that the
constraints (\ref{constn}) can be solved by writing
\cite{angelucci,css}
\begin{equation}
\vec n_{1} + i \vec n_{2 } \equiv \epsilon_{ab} w_{b} \vec
\sigma_{ac} w_c \label{z2gn}
\end{equation}
where $w_a$ is a 2 component complex spinor obeying $|w_\uparrow
|^2 + |w_\downarrow |^2 =1 $. It is useful to compare (\ref{z2gn})
with (\ref{nz}), which parameterized a single vector also in terms
of a complex spinor $z_a$. Whereas (\ref{nz}) was invariant under
the U(1) gauge transformation (\ref{gauge1}), notice that
(\ref{z2gn}) is only invariant under the $Z_2$ gauge
transformation
\begin{equation}
w_a (r, \tau) \rightarrow \varrho (r, \tau) w_a (r, \tau)
\label{z2g}
\end{equation}
where $\varrho (r, \tau) = \pm 1$ is an arbitrary field which
generates the gauge transformation. This $Z_2$ gauge
transformation will play an important role in understanding the
structure of the paramagnetic phase
\cite{ReSaSpN,ijmp,sst,Wen,sf}.

We can now study fluctuations of the non-collinear antiferromagnet
by expressing the effective action in terms of the $w_a$. Apart
from the familiar constraints of spin rotational invariance, and
those imposed by (\ref{z2g}), the effective action must also obey
the consequences of translational invariance which follow from
(\ref{neelt}); the action must be invariant under
\begin{equation}
w_a \rightarrow w_a e^{-i \vec Q \cdot \vec a/2}
\end{equation}
where $\vec a$ is any triangular lattice vector. In the continuum
limit, this leads to the following effective action
\begin{equation}
\mathcal{Z}_w = \int \mathcal{D} w_a (r, \tau) \exp \Biggl( - \int
d^2 r d \tau \biggl[ |\partial_\mu w_a |^2 + s |w_a |^2 +
\frac{u}{2} (|w_a|^2)^2 \biggl]\Biggl); \label{zw}
\end{equation}
notice there is a free integration over the $w_a$, and so we have
softened the rigid length constraint. Comparing this with
(\ref{zc}), we observe that the U(1) gauge field is now missing,
and we simply have a Landau-Ginzburg theory for a 2 component
complex scalar. The $Z_2$ gauge invariance (\ref{z2g}) plays no
role in this continuum critical theory for the destruction of
non-collinear magnetic order, but as we discuss below, it will
play an important role in the analysis of the paramagnetic phase.
The theory (\ref{zw}) has a global O(4) invariance of rotations in
the 4-dimensional space consisting of the real and imaginary parts
of the 2 components of $w_a$: consequently the critical exponents
of (\ref{zw}) are identical to those of the well known 4-component
$\varphi^4$ field theory. Notice that there is no O(4) invariance
in the microscopic theory, and this symmetry emerges only in the
continuum limit \cite{joliceour,css}: the simplest allowed term
which breaks this O(4) invariance is $|\epsilon_{a b} w_a
\partial_\mu w_b |^2$, and this term is easily seen to be
irrelevant at the critical point of the theory (\ref{zw}).

Let us now turn to a discussion of the nature of the paramagnetic
phase obtained in the region of large positive $s$ in (\ref{zw}).
Here, the elementary excitations are free $w_a$ quanta, and these
are evidently $S=1/2$ spinons. There is also a neutral, spinless
topological excitation \cite{rc,ReSaSpN,ijmp} whose importance was
stressed in Ref.~\cite{sf}: this is the `vison' which is
intimately linked with the $Z_2$ gauge symmetry (\ref{z2g}). It is
a point defect which carries $Z_2$ gauge flux. The vison has an
energy gap in the paramagnetic phase, and indeed across the
transition to the magnetically ordered state. This was actually
implicit in our taking the continuum limit to obtain the action
(\ref{zw}). We assumed that all important spin configurations
could be described by a smooth, single-valued field $w_a (r,
\tau)$, and this prohibits vison defects around which the $w_a$
are double-valued. It is also believed that the vison gap allows
neglect of Berry phase effects across the transition described by
(\ref{zw}): after duality, the Berry phases can be attached to
monopoles and visons \cite{sf,JS}, and these are suppressed in
both phases of Fig~\ref{fig13}.

\section{Conclusions}
\label{sec:conc}

This article has described a variety of quantum phases of
antiferromagnetic Mott insulators, and the transitions between
them.

Let us first summarize the phases obtained in zero applied
magnetic field, and transitions that can be tuned between them by
varying the ratio of exchange constants in the Hamiltonian
(experimentally, this can be achieved by applied pressure). The
magnetically ordered states discussed were the collinear N\'{e}el
state (shown in Figs~\ref{fig4} and~\ref{fig12}), and the
non-collinear `spiral' (shown in Fig~\ref{fig13}). We also found
paramagnetic states which preserved spin rotation invariance and
which had an energy gap to all excitations: these include the
dimerized states (shown in Figs~\ref{fig2} and~\ref{fig4}), the
related bond-ordered states which spontaneously break lattice
symmetries (shown in Figs~\ref{fig11} and~\ref{fig12}), and the
`resonating valence bond' paramagnet with topological order and
deconfined spinons (shown in Fig~\ref{fig13}). The continuous
quantum phase transitions we found between these states were:\\
({\em a\/}) the transition between the dimerized paramagnet and
the collinear N\'{e}el state (both states shown in Fig~\ref{fig4})
was described by the theory $\mathcal{S}_{\varphi}$ in (\ref{sp});\\
({\em b\/}) the transition between the dimerized paramagnet and a
non-collinear magnetically ordered state was described by
$\mathcal{S}_{\Phi}$ in (\ref{sp2});\\
({\em c\/}) the transition between the collinear N\'{e}el state
and the paramagnet with {\em spontaneous\/} bond order (shown in
Fig~\ref{fig12}) was described for
$S=1/2$ antiferromagnets by $\mathcal{Z}_c$ in (\ref{zc});\\
({\em d\/}) the transition between the state with non-collinear
magnetic order and the RVB paramagnet (both states shown in
Fig~\ref{fig13}) was described by $\mathcal{Z}_w$ in (\ref{zw}).

We also mention here other quantum transitions of Mott insulators,
which involve distinct paramagnets on {\em both\/} sides of the
critical point. These we did not discuss in the present paper, but
such transitions have been discussed in the literature:\\
({\em e\/}) the transition between a paramagnet with spontaneous
bond order (Fig~\ref{fig12}) and a RVB paramagnet
(Fig~\ref{fig13}) is described by a compact U(1) lattice gauge
theory with charge 2 Higgs fields (closely related to
$\mathcal{Z}_1$ in (\ref{z1})), and is discussed in
Refs.~\cite{JS,jpsj,glw};\\
({\em f\/}) transitions between paramagnets with different types
of spontaneous bond order can be mapped onto transitions between
different smooth phases of height models like (\ref{he1}), and are
discussed in Refs.~\cite{vbs,sondhivbs}.

Section~\ref{sec:mag} also considered quantum transitions that
could be tuned by an applied magnetic field. We mainly considered
the case of the coupled-dimer antiferromagnet, but very similar
theories apply to the other states discussed above. The general
theory has the structure of $\mathcal{S}_{\Psi}$ in (\ref{spsi}),
describing the Bose-Einstein condensation of the lowest non-zero
spin quasiparticle excitation of the paramagnet. For the coupled
dimer model this quasiparticle had $S=1$, but an essentially
identical theory would apply for cases with $S=1/2$ spinon
quasiparticles.

\subsection*{Acknowledgements}

This research was supported by the National Science Foundation
under grant DMR-0098226. I am grateful to Lorenz Bartosch,
Gregoire Misguich, Flavio Nogueira, T.~Senthil, and Asle Sudb\o~
for valuable comments, and T.~Senthil, Ashvin Vishwanath, Leon
Balents, and Matthew Fisher for a fruitful recent collaboration
\cite{senthil,glw}, upon which Section~\ref{sec:qpt} is based, on
the breakdown of Landau-Ginzburg-Wilson theory in Mott insulators.

%%%%%%%%%%%%%%%%%%%%%%%%%%%%%%%%%%%%%%%%%%%%%%%%%%%%%%%%%%%%%%%%%%%%%%  }

%%%%%%%%%%%%%%%%%%%%%%%%%%%%%%%%%%%%%%%%%%%%%%%%%%%%%%%%%%%%%%%%%%%%%%

\printindex

\begin{thebibliography}{99.}
\bibitem{slater} Insulators with an even number of electrons per
unit cell can be adiabatically connected to band insulators, and
so some readers may object to calling such materials `Mott
insulators'. However, the very different energy scales of the spin
and charge excitations in the experimental systems are best
understood in the framework of the Mott theory, and one regards
the dimerization as a small, low energy, deformation. Following
widely accepted practice, we will continue to label these
materials Mott insulators.
\end{thebibliography}
\end{document}